\documentclass[%
 reprint,
 amsmath,amssymb,
 aps,
 superscriptaddress,
]{revtex4-2}
\usepackage{siunitx}
\usepackage{bbm}
\usepackage{graphicx}
\usepackage[colorlinks=true, allcolors=blue]{hyperref}
\usepackage{placeins}

\graphicspath{{plots/}}

\title{Stacking domain morphology in epitaxial graphene on silicon carbide}
\date{\today}

\newcommand{\figref}[2][]{%
\ifx\FirstArg\empty
Figure \ref{fig:#2}%
\else
Figure \ref{fig:#2}\subref{#1}%
\fi}

\newcommand{\subref}[1]{\textbf{#1}}
\newcommand{\subf}[1]{\subref{#1}\textbf{,}}

\begin{document}
\title{Stacking domain morphology in epitaxial graphene on silicon carbide}
\author{Tobias A. de Jong}
\email{jongt@physics.leidenuniv.nl}
\affiliation{Huygens-Kamerlingh Onnes Laboratorium, Leiden Institute of Physics, Leiden University, Niels Bohrweg 2, P.O. Box 9504, NL-2300 RA Leiden, The Netherlands}
\author{Luuk Visser}
\affiliation{Huygens-Kamerlingh Onnes Laboratorium, Leiden Institute of Physics, Leiden University, Niels Bohrweg 2, P.O. Box 9504, NL-2300 RA Leiden, The Netherlands}
\author{Johannes Jobst}
\affiliation{Huygens-Kamerlingh Onnes Laboratorium, Leiden Institute of Physics, Leiden University, Niels Bohrweg 2, P.O. Box 9504, NL-2300 RA Leiden, The Netherlands}
\author{Ruud M. Tromp}
\affiliation{IBM T.J.Watson Research Center, 1101 Kitchawan Road, P.O.\ Box 218, Yorktown Heights, New York 10598, USA}
\affiliation{Huygens-Kamerlingh Onnes Laboratorium, Leiden Institute of Physics, Leiden University, Niels Bohrweg 2, P.O. Box 9504, NL-2300 RA Leiden, The Netherlands}
\author{Sense Jan van der Molen}
\email{molen@physics.leidenuniv.nl}
\affiliation{Huygens-Kamerlingh Onnes Laboratorium, Leiden Institute of Physics, Leiden University, Niels Bohrweg 2, P.O. Box 9504, NL-2300 RA Leiden, The Netherlands}
\begin{abstract}

Terrace-sized, single-orientation graphene can be grown on top of a carbon buffer layer on silicon carbide by thermal decomposition. Despite its homogeneous appearance, a surprisingly large variation in electron transport properties is observed.

Here, we employ Aberration-Corrected Low-Energy Electron Microscopy (AC-LEEM) to study a possible cause of this variability. We characterize the morphology of stacking domains between the graphene and the buffer layer of high-quality samples. Similar to the case of twisted bilayer graphene, the lattice mismatch between the graphene layer and the buffer layer at the growth temperature causes a moir\'e pattern with domain boundaries between AB and BA stackings.

We analyze this moir\'e pattern to characterize the relative strain and to count the number of edge dislocations. Furthermore, we show that epitaxial graphene on silicon carbide is close to a phase transition,
causing intrinsic disorder in the form of co-existence of anisotropic stripe domains and isotropic trigonal domains. 
Using adaptive geometric phase analysis, we determine the precise relative strain variation caused by these domains. We observe that the step edges of the SiC substrate influence the orientation of the domains and we discuss which aspects of the growth process influence these effects by comparing samples from different sources.
\end{abstract}
\maketitle
\section{Introduction}

Epitaxial graphene on silicon carbide can be grown on the wafer scale by thermal decomposition, on both doped and insulating SiC substrates. 
As silicon has a lower sublimation point than carbon, heating an atomically flat surface of SiC to 1200$^\circ$C or higher, the silicon sublimates, while the carbon stays behind~\cite{van_bommel_leed_1975,riedl_structural_2007}.
The first layer of carbon is still covalently bonded to the substrate. 
This means this so-called buffer layer is insulating as it lacks full sp$^2$ hybridization.
The subsequent layer(s) do exhibit full sp$^2$ hybridization and are therefore only bonded to the other layer(s) by Van der Waals forces.
Growing the graphene at higher temperatures or keeping it hot for longer causes more silicon to sublimate and extra layers to form between the buffer layer and the lowest graphene layer~\cite{tanaka_anisotropic_2010}. 
Although growth on the carbon face of the SiC is possible, we here focus on the more homogeneous graphene growth on the Si face. 
To create more regular layers, a gas backpressure of silane~\cite{tromp_thermodynamics_2009} or, more commonly, argon of up to one bar can be supplied. This achieves more uniform growth at lower speeds and higher temperatures~\cite{emtsev2009-towards}. Additionally, extra carbon can be provided by depositing carbon in advance~\cite{kruskopf_comeback_2016,momeni_pakdehi_minimum_2018}.
Optimization of these growth procedures has led to very homogeneous monolayer graphene on SiC samples, but a surprisingly large variation of electron transport properties remains in these samples~\cite{sonde_role_2010}.

The graphene lies on a buffer layer covalently bonded to the SiC and forms a $(6\sqrt3 \times 6\sqrt3)R30^\circ$ reconstruction with the underlying SiC lattice~\cite{riedl_structural_2007,kim_origin_2008}.
However, as the ratio of the lattice constants of graphene and SiC does not perfectly adhere to the ratio given by the reconstruction and both materials exhibit different thermal expansion rates, stacking domain boundaries occur to resolve the additional lattice mismatch. 
Such domains have been observed before, including using Dark Field Low Energy Electron Microscopy (LEEM), thermoelectric imaging, and Scanning Tunneling Microscopy~\cite{dejong2018intrinsic,cho_thermoelectric_2013,hibino_stacking_2009,lalmi_flower-shaped_2014}.

Here, we study these stacking domain boundaries in high quality epitaxial graphene obtained via three different growth processes.
We directly image the domain boundaries in these samples using Bright Field Low Energy Electron Microscopy at a landing energy $E_0 \approx 40$\,eV~\cite{de_jong_2022_contrast}.
By employing stitching of high resolution AC-LEEM data as described in~\cite{de_jong_imaging_2021}, we obtain a field of view exceeding $10\times10$\,$\mu$m$^2$, while retaining a high resolution of at least 2.2\,nm/pixel to characterize the stacking domain boundaries, enabling the gathering of statistics and the extraction of properties of the graphene itself.

\begin{figure*}[!ht]
\centering
\includegraphics[height=0.8\textheight]{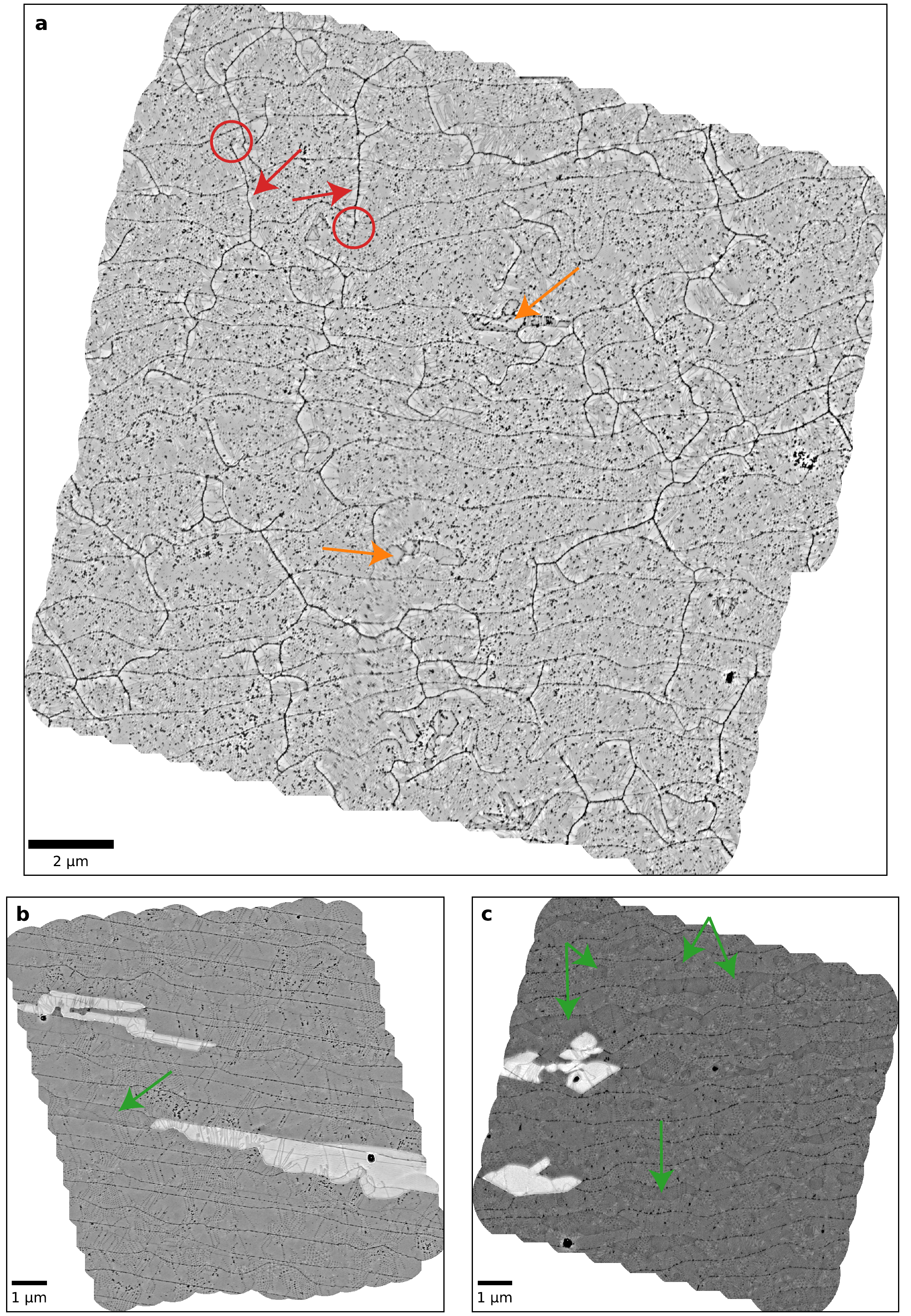}
\caption{\textbf{Sample overviews.} 
\subf{a} Sample A, intercalated quasi-freestanding graphene on SiC, grown using an experimental carbonated growth from Heiko Weber's group in Erlangen-N\"urnberg imaged using BF-LEEM at $E_0=40$\,eV.
\subf{b} Sample B, commercially bought graphene on 4H-SiC from Graphensic based on the Link\"oping growth technique, imaged at $E_0=37$\,eV.
\subf{c} Sample C, polymer assisted growth on 6H-SiC sample grown at PTB Braunschweig imaged at $E_0=36$\,eV.
All full areas are normalized by dividing by a Gaussian smoothed version of the image (with width $\sigma=50$ pixels) to eliminate global brightness variation, treating (brighter) bilayer areas separately for \subref{b,c}. Details about the indicated features are given in the main text.}
\label{fig:normalizedEGsampleoverviews}
\end{figure*}

Samples from three different sources are imaged in this way. 
First, \textit{sample A}, grown in the Weber group at the University of Erlangen-N\"urnberg using an early prototype of a variation on polymer assisted growth, where a layer of carbon is sputtered onto the SiC before growth. This sample is grown with an argon back pressure to enable uniform growth and hydrogen intercalated to create quasi-freestanding bilayer graphene~\cite{ott_light_2021}. Second, \textit{sample B}, a commercially bought sample from the company Graphensic, which bases its growth process on the work of the group of Professor Yakimova at Link\"oping University~\cite{yazdi_growth_2013}. Finally, \textit{sample C}, grown at the PTB in Braunschweig using polymer assisted growth in argon back pressure~\cite{kruskopf_comeback_2016,momeni_pakdehi_minimum_2018}.

In the next section the full datasets are shown and the visible features in the images are described qualitatively. Then, we will use GPA to quantitatively analyze the domain sizes and connect this to the relative strain between the layers. 
In Sections \ref{sec:spirals} and \ref{sec:Gedgedislocations} we will take a closer look at two peculiar features: spiral domain walls and edge dislocations, before we will interpret the results and draw conclusions about strain and local variation in these materials.

The Python code used to generate the figures in this work is available as open source at Ref.~\cite{graphene-stacking-domains-code}.

\section{Qualitative description of sample features}

Before analysis, the full datasets are normalized by dividing by a smoothed version of itself with a width of $\sigma=50$ pixels, where for samples B and C the bilayer areas are normalized separately.

For sample A an area of $305\mu \text{m}^2$ is imaged at a resolution of 2.2\,nm/pixel with an average total integration time of $194.7\,$s (for each pixel). 
For sample B an area of $111\mu \text{m}^2$, of which 7.1\% is bilayer, is imaged at 1.4\,nm/pixel with an average total integration time of $103.8\,$s. 
For sample C an area of $112\mu \text{m}^2$, of which 3.5\% is bilayer, is imaged at a resolution of 2.2\,nm/pixel with an average total integration time of $16.6\,$s.

\begin{figure}[!h]
\includegraphics[width=\columnwidth]{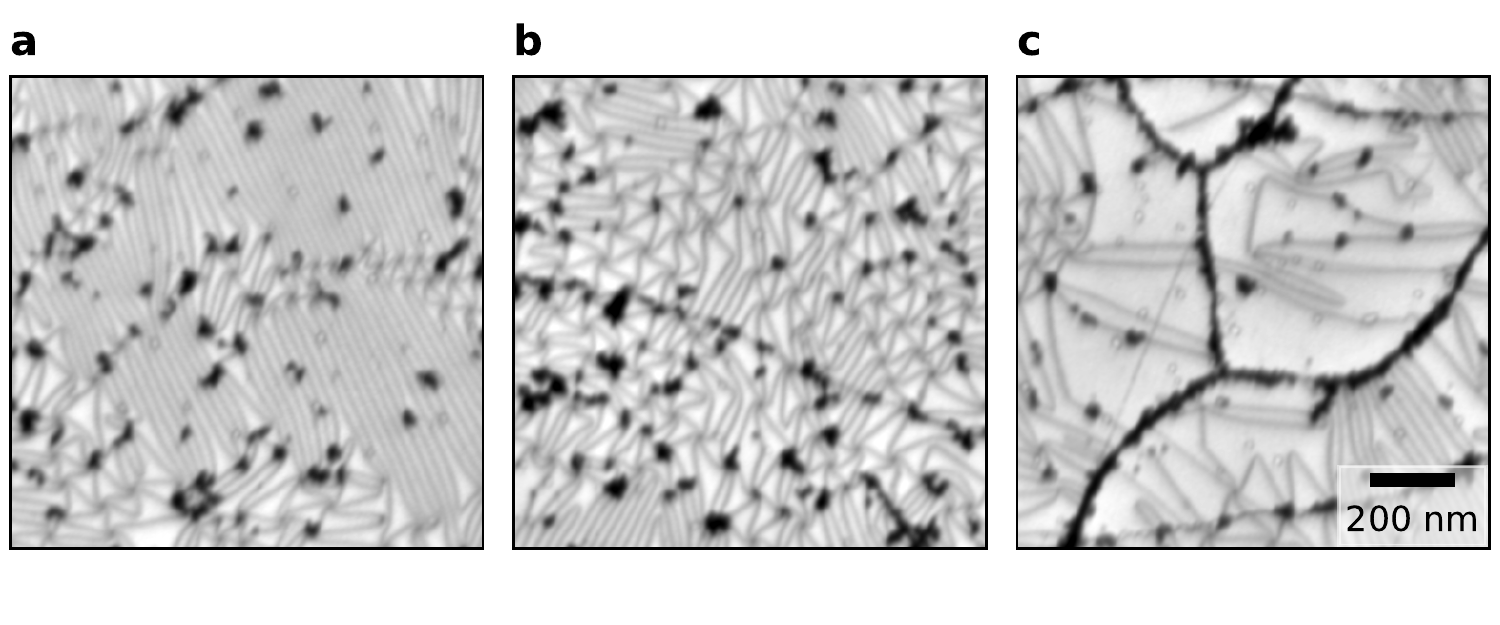}
\caption{\textbf{Details in sample A.} \subf{a} Striped stacking domains
\subf{b} Chaotic triangular stacking domains. Domain boundaries cross the horizontal step edge in this field of view.
\subf{c} Along the vertical dark features on this sample, significantly larger domains occur.
All panels have the same scale.}
\label{fig:CQFBLGdetails}
\end{figure}

For all three samples, terrace step edges of the SiC substrate are visible, running roughly horizontal. On all samples, some conglomerated carbohydrate adsorbates are visible as black spots, sticking to these substrate step edges and some other defects. 
Sample A shows more adsorbates than the other samples, but this is not due to the growth procedure, but due to sample handling and imaging conditions. 
This sample also shows additional defect lines running roughly vertical. Two examples are indicated with red arrows in \figref[a]{normalizedEGsampleoverviews}. As they terminate in points (indicated with red circles) and cross substrate steps, they seem unlikely to be terrace edges. Instead, they are probably folds or residue from excess carbon from the experimental carbonated growth process.

All three samples show some bilayer areas, occurring bright in \figref[b,c]{normalizedEGsampleoverviews}. For sample A, they occur dark (e.g. just above and below the center, indicated with orange arrows in \subref{a}), due to the different value of $E_0$ used and the hydrogen intercalation.
Furthermore both the samples in \figref[b,c]{normalizedEGsampleoverviews} show terraces with a slightly different contrast, next to the lower bilayer area in \subref{b} (indicated with a green arrow) and in several spots in \subref{c}, in particular in round spots in the center of terraces (some examples indicated with green arrows). This difference in intensity is due to the stacking order and termination of the underlying silicon carbide~\cite{momeni_pakdehi_silicon_2020,sinterhauf_substrate_2020}.

Of the three samples, Sample A is the most irregular. In addition to the aforementioned vertically running defect lines, the SiC substrate step edges are wavier than in the other two samples, although further apart, due to a step bunching procedure applied in the process before graphene growth as described in the methods section of Ref.~\cite{dejong2018intrinsic}.
In \figref{CQFBLGdetails}, some full-resolution detail images showcase the domain boundary morphology. The domain shapes are irregular. Stripe domains (\figref[a]{CQFBLGdetails}) occur in roughly three directions. Triangular domains occur as well, but are irregularly shaped, not forming larger regular grids (\figref[b]{CQFBLGdetails}). Remarkably, around the defect lines, domains are significantly larger and irregular (\figref[c]{CQFBLGdetails}), suggesting they are folds out of plane which absorb some of the lattice mismatch.

\begin{figure}[!ht]
\includegraphics[width=\columnwidth]{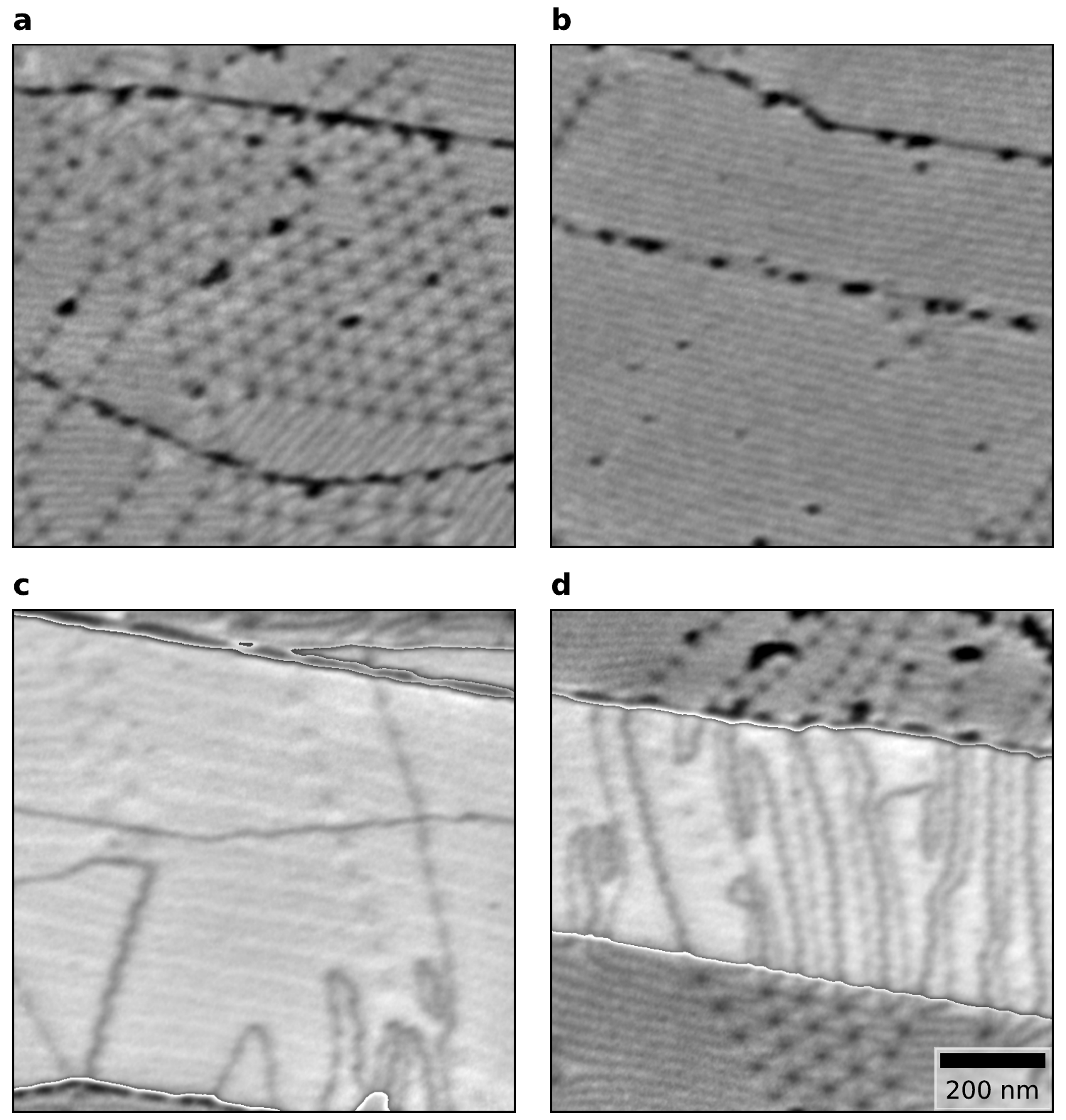}
\caption{\textbf{Details in sample B.} \subf{a} More regular triangular stacking domains.
\subf{b} Very large and regular striped domains with high stripe density.
\subf{c} Domains in the bilayer region: Twinned stripe domains between the buffer layer and the lower graphene layer (faint horizontal lines) and irregular domains between the two graphene layers.
\subf{d} Relation between triangular domains in the monolayer and domain boundaries in the bilayer area.
All panels have the same scale.}
\label{fig:linkopingdetails}
\end{figure}

Sample B is much more regular. 
In \figref[a,b]{linkopingdetails}, it is visible that both relatively regular triangular domains and very dense stripe domains occur~\cite{speck2017growth,butz2014dislocations}. 
Although these stripe domains again occur in three directions, they occur parallel to the substrate step edges in the vast majority of the cases.
Details of the bilayer-on-buffer layer areas are shown in \figref[c,d]{linkopingdetails}. Stripe domain boundaries occur between the buffer layer and the lower graphene layer, with the domain boundaries `twinning', i.e. forming pairs closer together, due to the energy mismatch between ABC and ABA stacked graphene~\cite{halbertal_moire_2021,guerrero-aviles_relative_2021,ravnik_strain-induced_2019}.
Domain boundaries between the top two graphene layers also occur, distinguishable by much higher contrast than those lower down, as expected. However, they seem largely irregular, which matches earlier observations that those domain boundaries are caused by nucleation instead of strain~\cite{dejong2018intrinsic,cho_thermoelectric_2013,hibino2009stacking}.
However, in some areas where triangular domains border the bilayer, e.g. in \figref[d]{linkopingdetails}, it seems that the domain boundaries in the monolayer on buffer layer connect to domain boundaries in the bilayer-on-buffer layer on both levels, i.e. alternating between the buffer layer and the bottom graphene layer and between the two graphene layers.
 
\begin{figure}[!ht]
\includegraphics[width=\columnwidth]{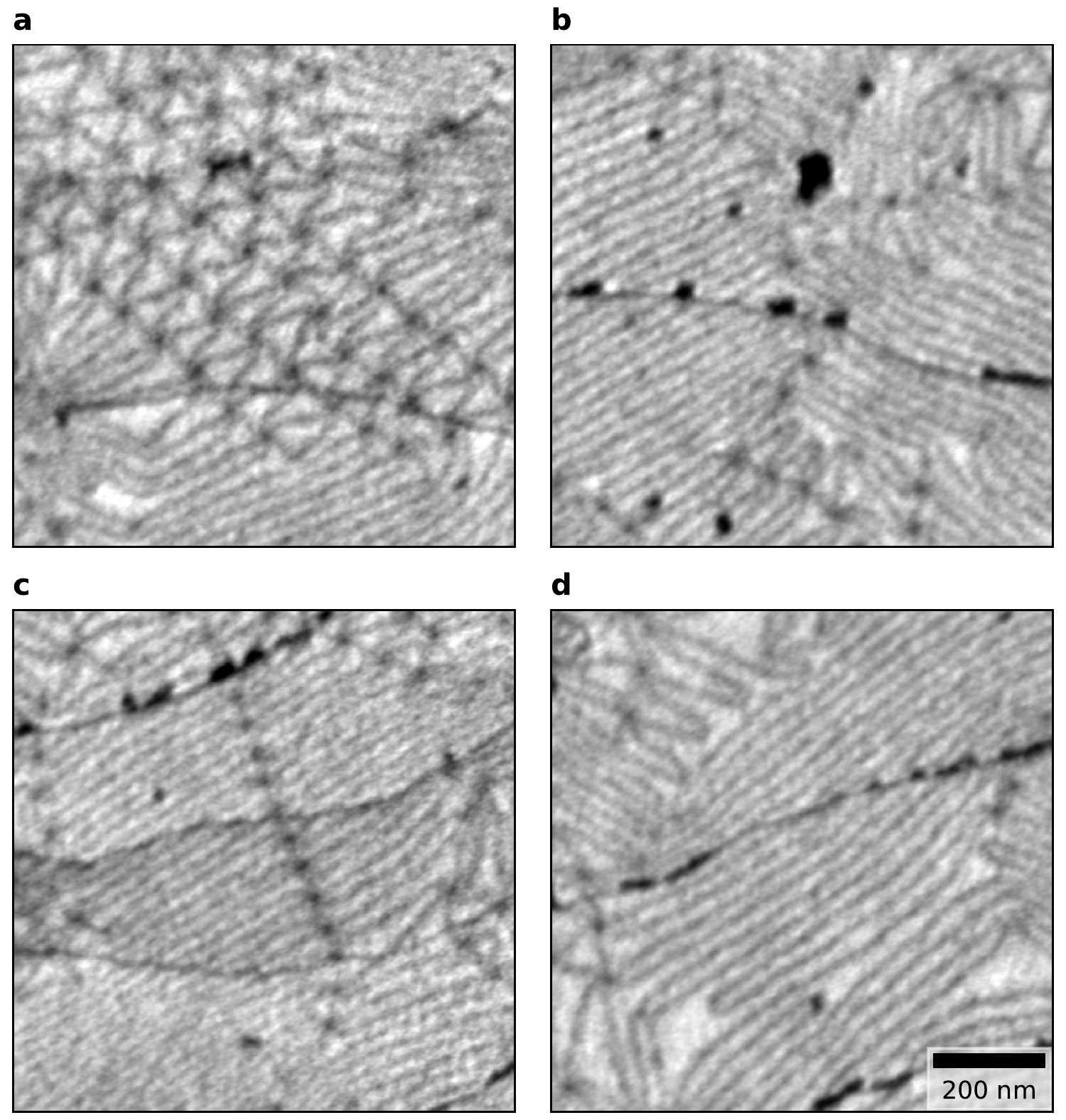}
\caption{\textbf{Details in sample C.} \subf{a} Large triangular domains.
\subf{b} Stripe domains in two distinct directions.
\subf{c} Stripe domains with a node line crossing across substrate step edges.
\subf{d} Low density stripes and disorder.
All panels have the same scale.}
\label{fig:braunschweigdetails}
\end{figure}

For Sample C, details are shown in \figref{braunschweigdetails}. Some triangular domains occur, but a larger part is covered by stripe domains in three directions. Both the triangles and stripe widths vary, but generally are significantly larger than in Sample B. Domain boundaries and even AA node strings (\figref[c]{braunschweigdetails}, also occurring on Sample B) seem to cross substrate step edges unperturbed. Finally, like in Sample A, irregular domain shapes are quite common and will be explored in Section \ref{sec:Gedgedislocations}.
But first, in the next section we will apply Geometric Phase Analysis (GPA) to quantify the domain morphology and leverage the large size of the imaged areas to obtain some statistics.

\section{Stripe domains in epitaxial graphene}
From the morphology of the domains as shown in the previous section, it was already clear that the graphene on SiC samples are less homogeneous than widely believed. 
Although strain in graphene on SiC and even non-homogeneity of the strain has been studied extensively using e.g. Raman spectroscopy~\cite{schmidt_strain_2011,schumann2014effect,ferralis_evidence_2008}, the mere existence of domain boundaries which concentrate the strain means that extra care should be taken interpreting these results, as these techniques average over relatively large areas.
In this section we will extract more quantitative information from the stacking domains, and use the relation between the stacking domains and the atomic lattice to quantify strain and disorder on the atomic level.

Assuming the amount of carbon atoms in the graphene layer is fixed after the growth stops, the average size of the domains is determined by the remaining mismatch between graphene and the $6\sqrt{3}$ reconstruction of the buffer layer on the SiC at the growth temperature.
The remaining mismatch is given by:
\[\epsilon = 1 - \frac{13a_\text{G} }{6\sqrt3 a_\text{SiC}}\]
Because the mismatch is relatively small (otherwise the reconstruction would not be able to form!), accurate values of the relevant lattice parameters and their temperature dependence are needed to calculate the expected domain size.
To obtain an estimate, we use the same values as used in Ref.~\cite{bao_synthesis_2016}: $a_\text{SiC} = 0.3096\,\text{nm}$ and $a_\text{G} = 0.2458\,\text{nm}$ at $T\approx 1200^\circ\text{C}$. This corresponds to a remaining lattice mismatch of $\epsilon = 0.7\%$ (where graphene has the smaller lattice constant compared to the buffer layer). Note that given the thermal expansion coefficients this number is strongly dependent on the growth temperature, decreasing by about $0.05\%$ for a 100K lower growth temperature.
Finally, it is claimed that a shorter growth time can give a small carbon deficiency, effectively yielding a tensile strain in the graphene layer at the growth temperature.

The two-dimensional Frenkel-Kontorova model can be used to describe domain forming in bilayer graphene. It predicts an extra phase transition if crossings of domain boundaries, i.e. AA-sites, cost extra energy compared to a non-crossing domain boundary. It is beneficial to form parallel domain boundaries instead of a triangular domain pattern by elongating triangular domains along one direction to essentially infinite length in one direction (See \figref[a]{GPAextraction})~\cite{lebedeva_two_2020}. Note that this even holds for samples under bilateral symmetric/isotropic strain, if the energy cost of AA-sites is high enough.
This is a discontinuous, symmetry-breaking transition and thus a first order phase transition.
For bilayer graphene, the ratio between stacking energy costs of AA stacking and SP stacking is about 9. This corresponds to stripe domains forming if the strain is above a lower critical strain value of $\epsilon_{c1} = 0.37\%$.
Therefore, stripe domains can be expected in graphene on SiC samples (assuming the graphene--buffer layer interaction is close enough to the one of bilayer graphene, or intercalated samples). For samples created with short enough growth times or at low enough temperature, a mixture of both parallel domain boundaries and triangular patterns can be expected.

Finally, note that as the sample cools after growth, the lattice mismatch between the graphene and the SiC reconstruction decreases (to 0.1\% or less at room temperature), but the number of carbon atoms in the graphene layer is already roughly fixed and the graphene layer is pinned to the substrate by defects and step edges.
This yields a total compressive strain on the graphene (which might be partially offset by a carbon deficiency tensile strain for short growth times), but the relative lattice mismatch is globally kept the same by defect pinning.

Indeed, this is what we observe for the graphene on SiC samples, where the periodicity of the buffer layer is forced by the underlying SiC substrate, but the behavior of the graphene layer on top of that is governed by the Van der Waals interaction and graphene's properties.

Note however, that unlike on some metals~\cite{hattab_interplay_2012,sutter_graphene_2009} for these boundaries all the strain compensation happens in-plane, i.e. no wrinkles form.

\subsection{Geometric phase analysis analysis of strain}

\begin{figure}[!ht]
\includegraphics[width=\columnwidth]{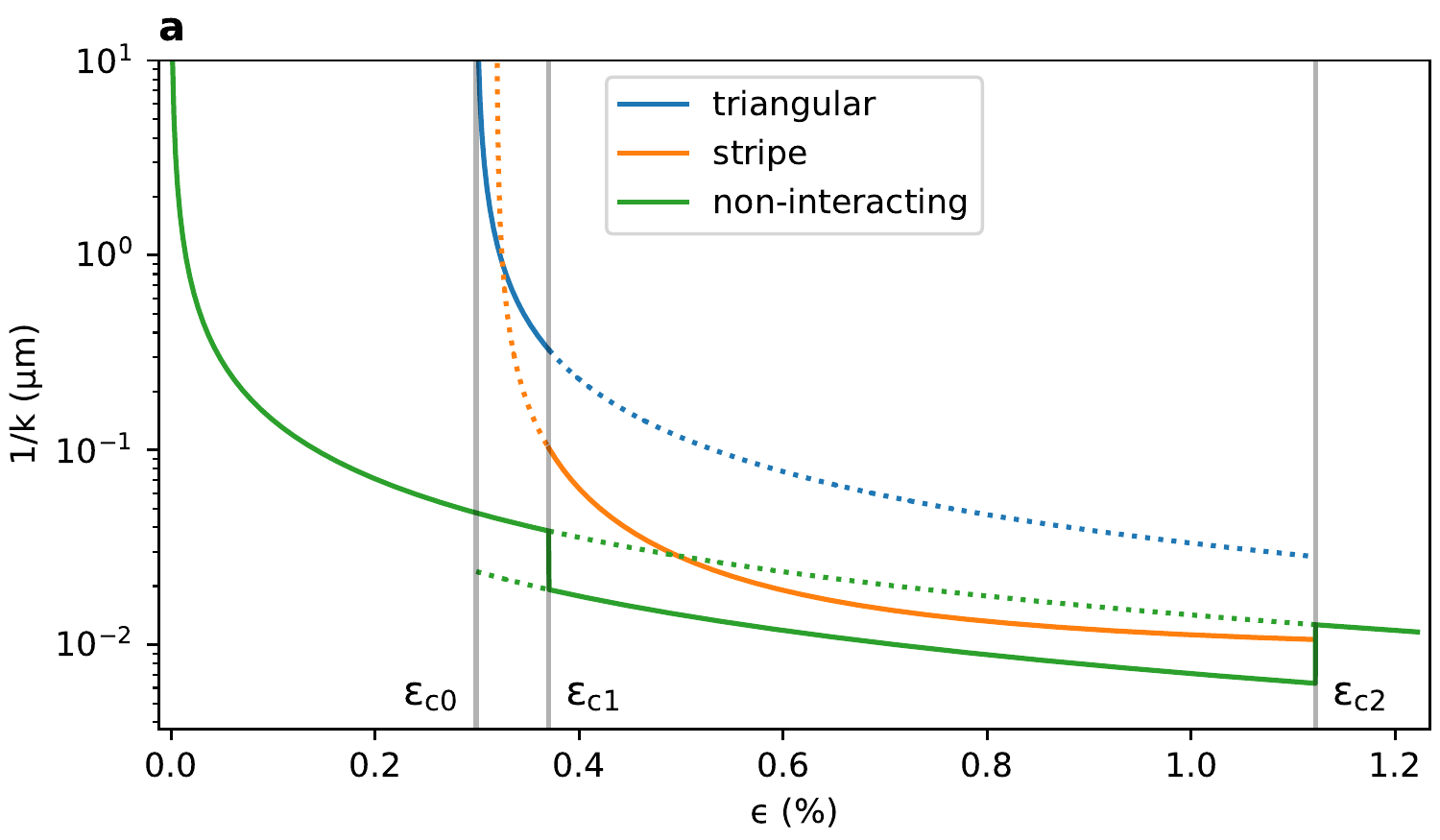}
\includegraphics[width=\columnwidth]{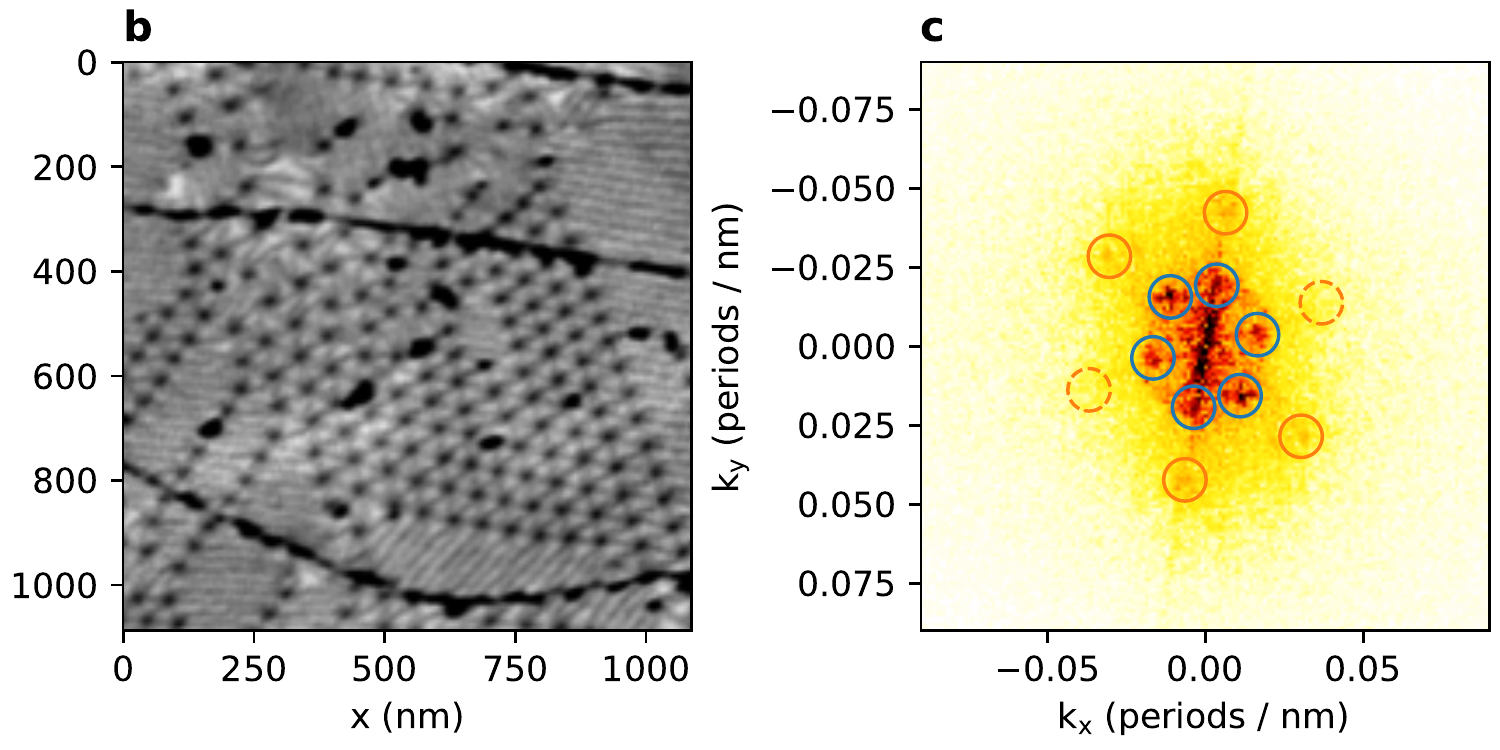}
\caption{\textbf{Stripe and triangular phases. } 
\subf{a} Calculated phase diagram and period of the superstructure as a function of relative isotropic strain in bilayer graphene as computed by Lebedeva and Popov for a barrier height of SP stacking of 1.61 meV/atom~\cite{lebedeva_two_2020}. 
Phases are from left to right: Commensurate, Triangular incommensurate, Striped incommensurate and again Triangular commensurate. The solid line indicates the observable line spacing in the superstructure in the respective phases. Also indicated is the periodicity for the non-interacting case, corresponding to the resulting average lattice mismatch.
\subf{b} Small crop of sample B with both triangular domains and stripe domain in two directions. A straight and a curved step edge run horizontally through the image and are decorated with adsorbates appearing in black.
\subf{c} Center of the FFT of \subref{b}. Detected triangular domain spots are circled in blue. Four detected stripe domain spots are circled orange, and their difference used as the third reference vector is circled in dashed orange.}\label{fig:GPAextraction}
\end{figure}

To characterize the stacking domains, we use (adaptive) geometric phase analysis, which uses a comparison to a perfect lattice to calculate the deformation of the domains~\cite{benschop_measuring_2021,de_jong_imaging_2021}. In this way, we can extract local periodicities from the real space images and calculate back to relative strain values.

The transition from triangular to striped domains causes the length of the corresponding $k$-vector to double as one direction of domain boundaries aligns with a second to become parallel, thus doubling the frequency (with the third being pushed out). Therefore stripe domains yield separate peaks at roughly double the frequency in the FFT of domain images compared to triangular domains, as highlighted in orange and blue respectively for Sample B in \figref{GPAextraction}.
This relatively large separation in Fourier space between the triangular phase and the three striped phases means we can perform GPA for each separately and use this to distinguish them on a large scale and characterize each phase independently.

\begin{figure*}[!ht]
\includegraphics[width=\textwidth]{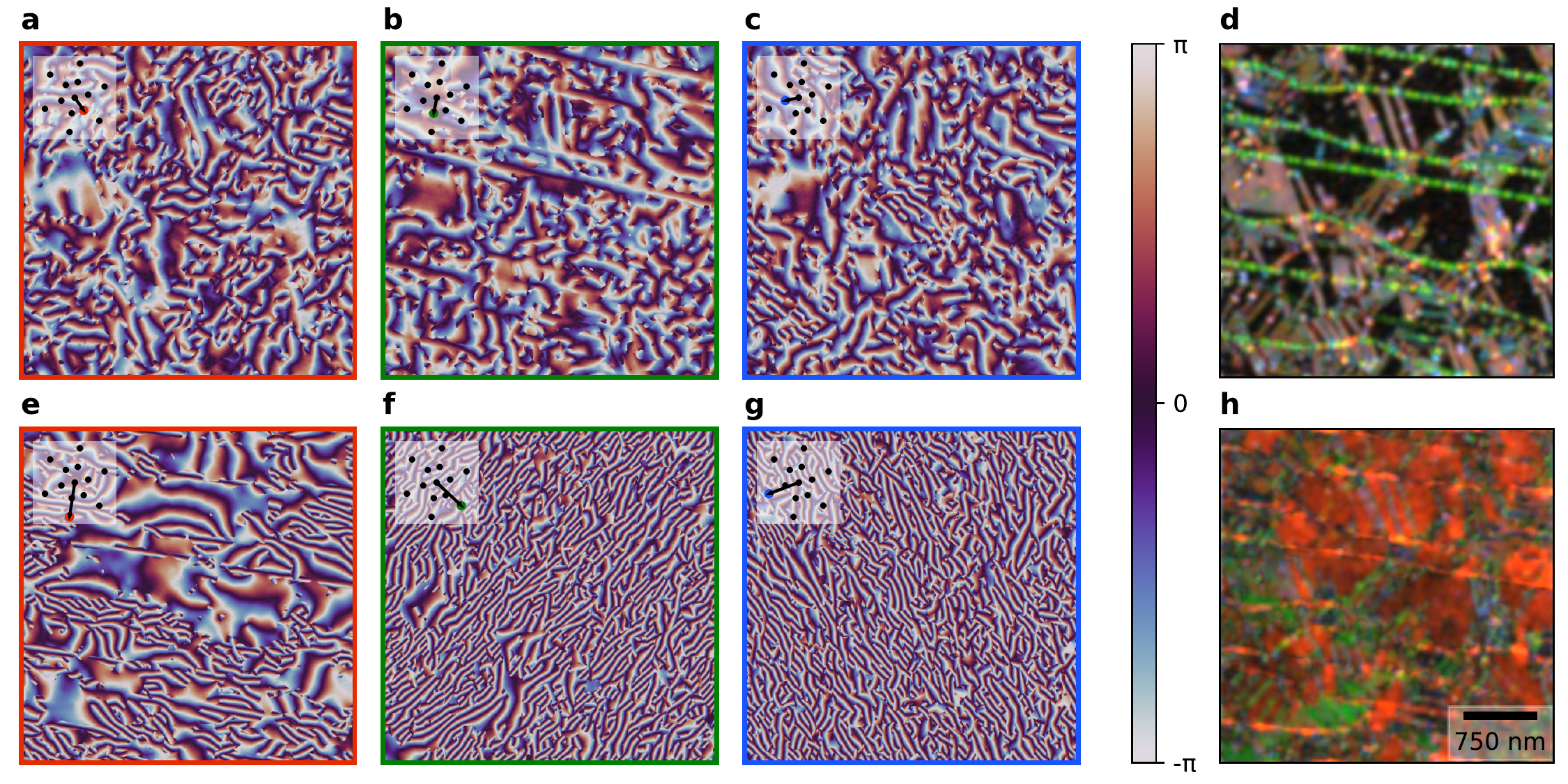}
\caption{\textbf{GPA phase analysis on an area of sample B with both triangular and stripe domains.} 
\subf{a-c} GPA phases corresponding to triangular domains. The $k$-vector used for each panel is indicated its inset.  
\subf{d} Normalized GPA amplitudes in RGB channels corresponding to the triangular domains. The substrate steps are visible as green lines.
\subf{e-g} GPA phases of stripe domains. Red (i.e. \subref{e}) corresponds to the dominantly present stripe phase.
\subf{h} Similar to \subref{d}, but for the stripe phases.}
\label{fig:GPAphasessampleB}
\end{figure*}

The GPA phases for the triangular phase of Sample B are shown in \figref[a-c]{GPAphasessampleB}, those for the stripe phase in \figref[e-g]{GPAphasessampleB}. For both the triangular and the stripe phase, the corresponding amplitudes are shown as red, green, and blue channels in Figure \ref{fig:GPAphasessampleB}\subref{d} and \subref{h} respectively.

The GPA phases only contain relevant information in the areas where the corresponding (stacking domain) phase occurs, as indicated by a high amplitude, and corresponding to a slowly varying GPA phase in real space.

\begin{figure}[!ht]
\includegraphics[width=\columnwidth]{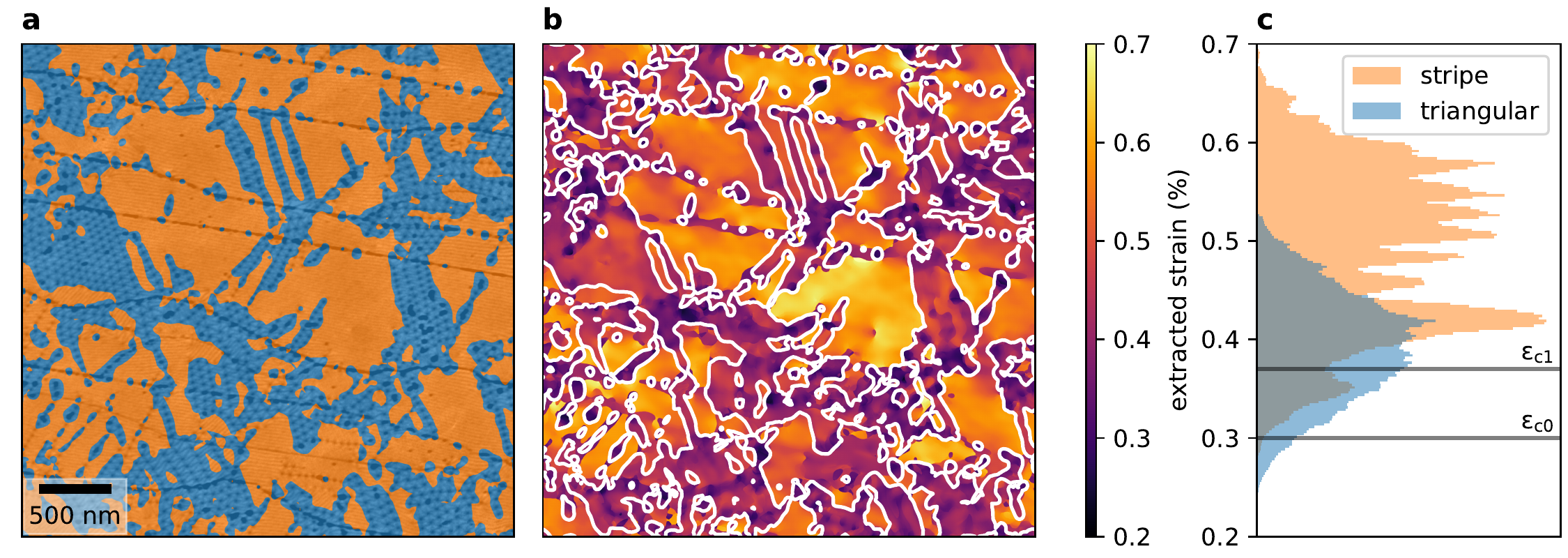}
\caption{\subf{a} Mask distinguishing stripe and triangular domains based on the triangle GPA magnitudes of sample B shown in \figref{GPAphasessampleB}. Stripe domains are shown in orange, triangles in blue.
\subf{b} Extracted strain for both types of domains in sample B for the same region. Boundary between the different domains is indicated with white lines.
\subf{c} Histogram of extracted strain values for both phases.}\label{fig:GPAstrainhistB}
\end{figure}

By comparing the GPA amplitudes, we create a mask dividing the sample in stripe domains and triangular domains.
For sample B, we use a threshold on the red and blue triangular domains, as the green GPA amplitude is dominated by the substrate steps, resulting in the mask shown in \figref[a]{GPAstrainhistB}, with 45\% of the characterized area triangular phase and 55\% stripe phase. The stripe phase is subdivided over the directions in 48\% red (parallel to the step edges), 6\% green, and less than 1\% blue, i.e. 88\% of the stripe domains is roughly parallel to the step edges.

For both the stripe domains and the triangular domains, we compute a local periodicity from the gradient of the GPA phases. This local periodicity we then convert the local relative strain between the layers, which is shown in \figref[b,c]{GPAstrainhistB}. Here, we have taken an average over the three directions for the triangular phase.
In total, we observe strain between 0.2\% and 0.7\%. On average, the stripe domains exhibit higher strain values than the triangular domains. Nevertheless, there is a large overlap and additionally a large part of the triangular domains exhibit a strain larger than the critical value $\epsilon_{c1}$. 

For sample A and C, the triangular domains were not regular enough to obtain a GPA signal. Nevertheless, The GPA amplitudes of the stripe domains indicate the stripe domains well. 
Therefore, as an alternative approach, masks are created by using a threshold value on these stripe domain amplitudes. Contrary to sample B, for sample A the three different stripe domains are almost divided equally, making up 18\%, 17\% and 19\% of the area respectively, for a total of 53\% (discrepancy due to rounding) stripe domains and 47\% otherwise.
Sample C is in between, with stripe domains making up 40\%, 11\% , and 10\% respectively for 61\% stripe domains in total. Here, like in sample B, the majority stripe direction (in red) is roughly parallel to the step edges.

\begin{figure}[!ht]
\includegraphics[width=\columnwidth]{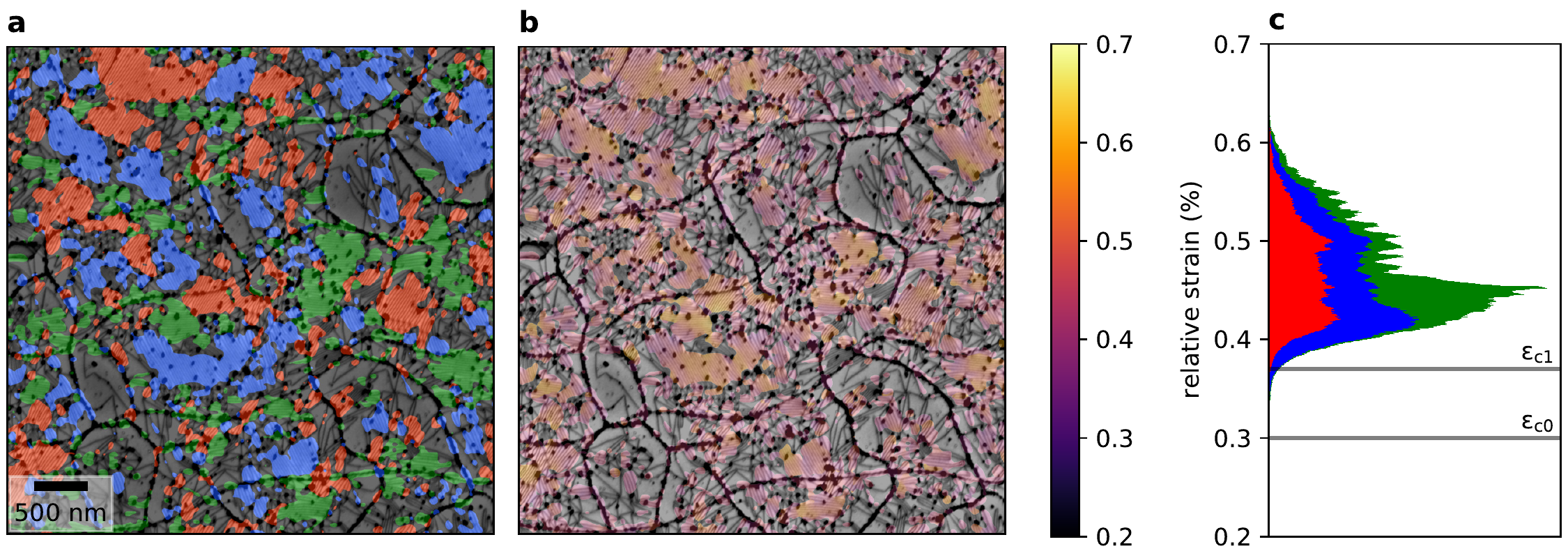}
\caption{\subf{a} Mask on sample A based on the GPA amplitudes labeling stripe domains in three directions in red, green and blue, showing stripe domains of several hundred nanometer across.
\subf{b} Extracted strain for all three directions of stripe domains for a region in the sample A.
\subf{c} Histogram of extracted strain values for the three stripe domain directions.}\label{fig:GPAstrainhistA}
\end{figure}

\begin{figure}[!ht]
\includegraphics[width=\columnwidth]{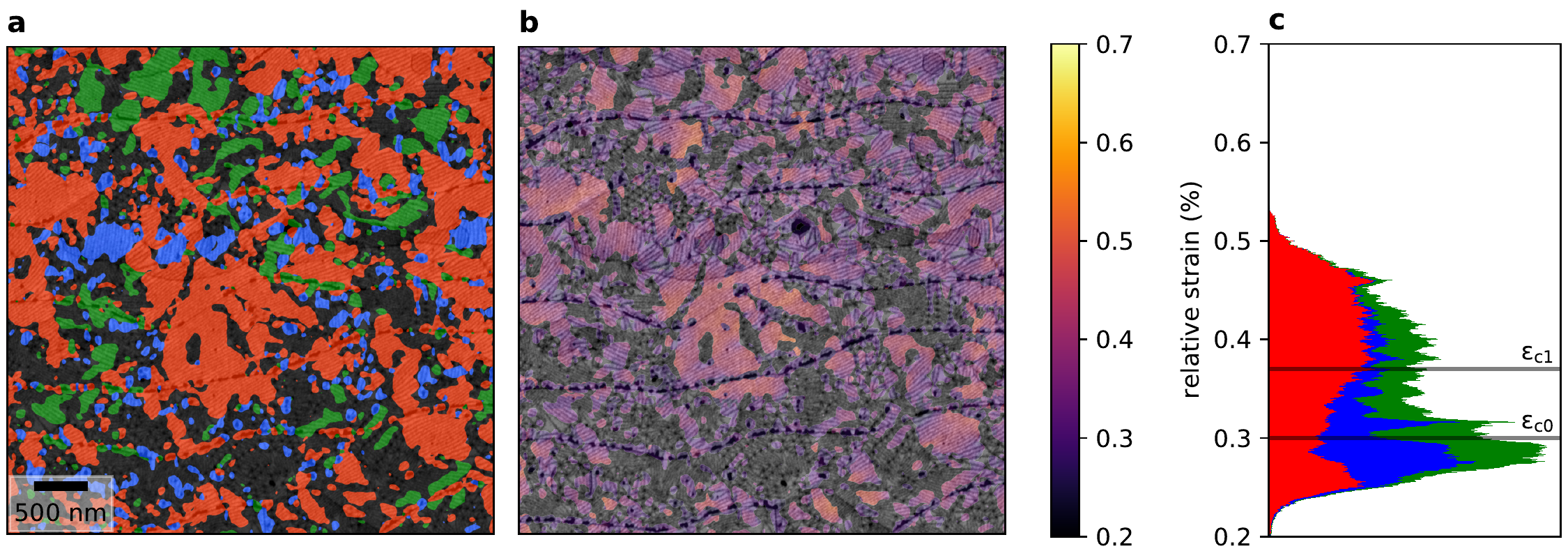}
\caption{\subf{a} Mask on sample C based on the GPA amplitudes labeling stripe domain in three directions in red, green and blue.
\subf{b} Extracted strain for all three directions of stripe domains for a region in the sample C.
\subf{c} Histogram of extracted strain values for the three stripe domain directions.}\label{fig:GPAstrainhistC}
\end{figure}

In \figref{GPAstrainhistA} and \figref{GPAstrainhistC} the occurrence of the three orientations of stripe domains and the extracted relative strain for these samples is shown. 
Notably, the extracted strain values for the three different samples cover different ranges, with the strain in the stripe domains in sample C significantly lower than in the others.

\subsection{Discussion of extracted strain}

Even allowing for error in the extraction of the strain, it is clear that a significant spread of the strain within a phase and a difference between the average strain value for the two phases exists. Furthermore, triangular domains occur at much higher relative lattice mismatch, i.e. the triangles are much smaller, than predicted by the Frenkel-Kontorova model, even when taking into account the error margins in the parameters of the model~\cite{lebedeva_two_2020}.

To interpret the stripe domains in terms of strain of the graphene, we need to closely consider what happens in such a stripe domain. Perpendicular to the stripe, all the lattice mismatch (at growth temperature) relative to the buffer layer is released and concentrated in the domain boundaries. Parallel to the stripe, we can roughly expect the same lattice mismatch, but now the graphene in that direction is strained to be commensurate to the buffer layer.
The variation in period and lattice mismatch perpendicular to the stripe is therefore also a measure of the variation and value of the strain itself parallel to the stripe.

Of course, the patches of anisotropic strain in different directions fit together, meaning that there will be local variation in the magnitude and direction of the strain, both relative to the substrate and in absolute terms.

\subsubsection{Domain wall orientations}
For sample B, domain walls in the stripe phases align with one of the domain wall directions in the triangular phase, i.e. the peaks in the FFT in \figref[c]{GPAextraction} for both phases are in the same direction. 
This is in contradiction with the theory of a strained lattice, where the domain walls in the triangular phase run along the zigzag directions of the graphene lattice and the domain walls in the stripe phases along one of the armchair directions~\cite{lebedeva_two_2020}.
Although the triangular domains in sample A and sample C are not ordered enough to show up as sharp peaks in the FFT, visual inspection of the images indicates a better adherence to this theoretical prediction, but also a spread of the orientation of the triangular domains which will be explored in more detail in the next section.
This does however suggest that the step edges in sample B have a strong influence on the direction of the domain boundary, presumably by uniaxially straining the lattice.

\section{Symmetry breaking AA-sites (Spiral domain walls)}\label{sec:spirals}

\begin{figure}[!ht]
\includegraphics[width=\columnwidth]{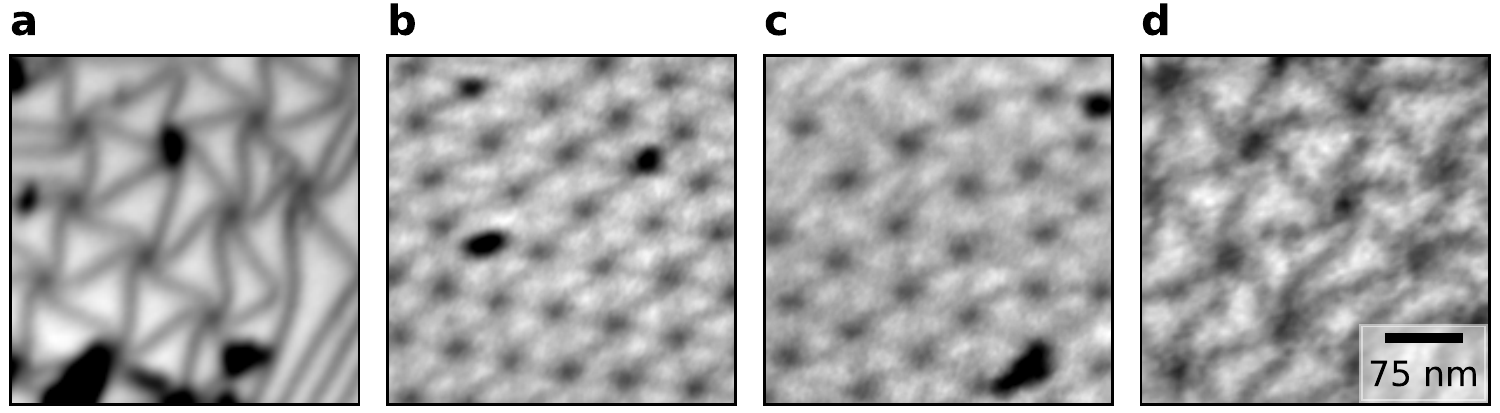}
\caption{ \subf{a} Spiral domain walls around AA-sites in sample A. 
\subf{b} Very regular triangles with very little spiral rotation in Sample B.
\subf{c} Less regular area of Sample B with more rotation in each spiral and opposing orientations.
\subf{d} Spiral domain walls in Sample C.
Contrast optimized per sample, scalebar valid for all images in this figure.}\label{fig:spiraldetails}
\end{figure}

We take a closer look at triangular domains in the different samples in \figref{spiraldetails}.
For these triangular domains, domain boundaries cross in nodes with six connecting domain walls, where the center of the node corresponds to AA stacking, therefore the nodes are labeled `AA-sites'.
Taking a closer look at the AA-sites, which appear as dark spots in \figref{spiraldetails}, we observe spiral domain walls. As the six domain boundaries approach a AA-site, they do not connect straight to it, but bend, either all to the left or all to the right, before connecting in a small spiral.
Such spiral domain boundaries have been observed before in various systems, including epitaxial metal systems such as Cu(111)/Ni(111) and Cu/Ru(1000) interfaces, graphene grown on copper, and 2H--1T polytype heterostructures in TaS$_2$~\cite{ravnik_strain-induced_2019,gutierrez_imaging_2016,shao_spiral_2013,hamilton_misfit_1995,gunther_strain_1995, lalmi_flower-shaped_2014} and have been reproduced in simulations~\cite{snyman_computed_1981,quan_tunable_2018}.

A tentative intuitive explanation for the occurance of these spiral domain walls is that they are a result of the shear domain boundary having lower energy cost per unit length than the strain type domain boundary. Thus a deviation from straight domain boundaries is promoted in the case of strain domain boundaries, but no such deviation is forced in the case of pure twist, where the domain boundaries are already in the lowest possible energy configuration~\cite{lebedeva_two_2020}.

In the samples studied here, both orientations of spirals occur, even on the same SiC terrace, even as direct neighbors.
According to simulations of a wide variety of twist angles, biaxial strain and combinations thereof~\cite{quan_tunable_2018}, a coexistence of both spiral orientations is an indication of pure lattice mismatch, without twist angle (as we would expect for this system), as the system is mirror symmetric and only the spiral itself breaks the symmetry.
A pure twist moir\'e pattern would be signaled by no spiraling and indeed no spiraling is visible in twisted bilayer graphene (TBG) samples~\cite{de_jong_imaging_2021,Yoo2019,verbakel_valley-protected_2021}. 
For a combination of strain and a small twist between the layers, the mirror symmetry is broken, and all spirals should align.

There is variation in how much the spirals curl near the AA sites.
It seems to depend on the sample, but there is even variation on the same sample, as exemplified by the difference between \figref[b,c]{spiraldetails}, both on Sample B.
Curiously, the moir\'e lattice also seems somewhat rotated between \figref{spiraldetails}b and c, indicating that in at least one of the two cases a local twist between the graphene and buffer layer occurs in addition to the biaxial strain.
A biaxial strain magnifies the relative twist of the atomic lattices, similar to how a twist angle magnifies a uniaxial strain. In this case, the atomic twist angle $\theta_a$ can be expressed as function of the biaxial strain $\epsilon$ and the apparent moir\'e twist angle $\theta$ as follows\footnote{For any realistic atomic strain, this expression is valid for $|\theta_a| \lesssim \epsilon$, but the crossover at $\theta = 30^\circ$ makes it impossible to distinguish the sign for $|\theta_a| \gtrsim \frac{\epsilon}{2}$ in BF-LEEM.}:
\[\theta_a = \theta - \arcsin\left(\frac{\sin\theta}{1+\epsilon} \right)\approx \epsilon \theta\]
Therefore, the observed moir\'e angle difference of the moir\'e patterns in \figref{spiraldetails}\subref{b} and \subref{c} of $\theta\approx 15^\circ$ at a strain of $\epsilon\approx0.45\%$ corresponds to a twist angle difference of the atomic lattices of $\theta_a \approx 0.07^\circ$. Here, the lattice in \subref{b} should actually correspond to a larger twist than the one in \subref{c}.
Note that this twist angle is the average angle between the unrelaxed lattices, as the atomic lattices within the domain are commensurate.

To fully analyze this, an optimization approach disentangling strain and twist just like employed for the case of twisted bilayer graphene could be used~\cite{de_jong_imaging_2021,Kerelsky2019}.

Nevertheless, this demonstrates that the triangular moir\'e pattern is not only a very sensitive measure of the lattice constant mismatch, but in the presence of such a lattice constant mismatch, the direction of the spiraling of the domain boundaries combined with the orientation of the moir\'e lattice is a very sensitive probe to the relative local orientation.

\section{Topological defects: Edge dislocations}\label{sec:Gedgedislocations}

The strain-caused moir\'e patterns observed in these systems magnify topological defects of the atomic lattice, just like in the twisted bilayer graphene case~\cite{de_jong_imaging_2021}.
More precisely, in the absence of any edge dislocations in the graphene layer and the buffer layer, each domain borders on precisely 3 AA-nodes and each pair of neighboring AA-nodes has only 1 domain boundary connecting them. Any deviation from these two rules indicates an atomic edge dislocation in one of the constituting layers.

\begin{figure}[h]
\includegraphics[width=\columnwidth]{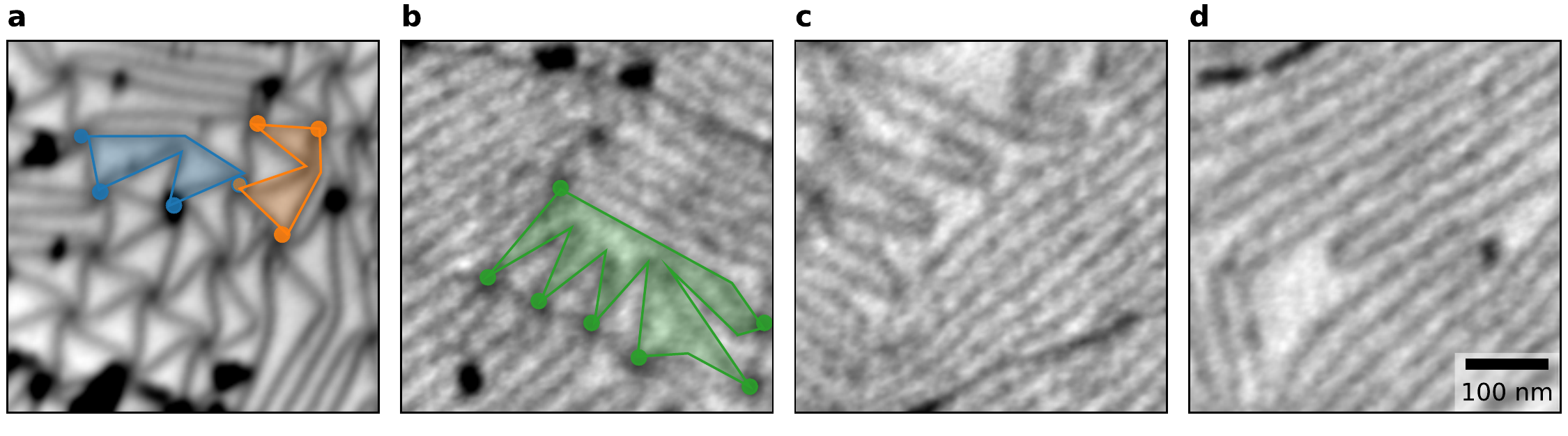}
\caption{ \subf{a} Detail of Sample A. Two sets of four AA-nodes, each set bordering one domain are indicated in blue and orange. Each of these domains therefore contains one edge dislocation.
\subf{b} Detail of Sample C. A set of seven AA-nodes bordering a single domain is indicated in green. This domain therefore contains multiple edge dislocations. Each edge dislocation also corresponds to a domain boundary with a characteristic kink in it, forming a triangle with another domain boundary connecting the same nodes.
\subf{c,d} Details of Sample C showing more kinked domain boundaries at the edges of striped domain areas.
Contrast optimized per sample, scalebar valid for all images in this figure.}\label{fig:dislocationdetails}
\end{figure}

Despite the fact that we can expect the number of edge dislocations in the underlying SiC substrate, and therefore in the buffer layer, to be very low on these microscopic levels~\cite{kimoto_defect_2020,lazewski_dft_2019}, we observe many such defects in all three samples, although in different densities. Examples for sample A and sample C are shown in \figref{dislocationdetails}.
Contrary to the TBG case, domain boundaries near edge dislocations in areas with more disorder deform the surrounding lattice significantly, with a domain boundary crossing over to the next domain boundary often with a significant kink, i.e. domain boundaries running between the same nodes repel each other and form a triangle.

\begin{figure}[!ht]
\includegraphics[width=\columnwidth]{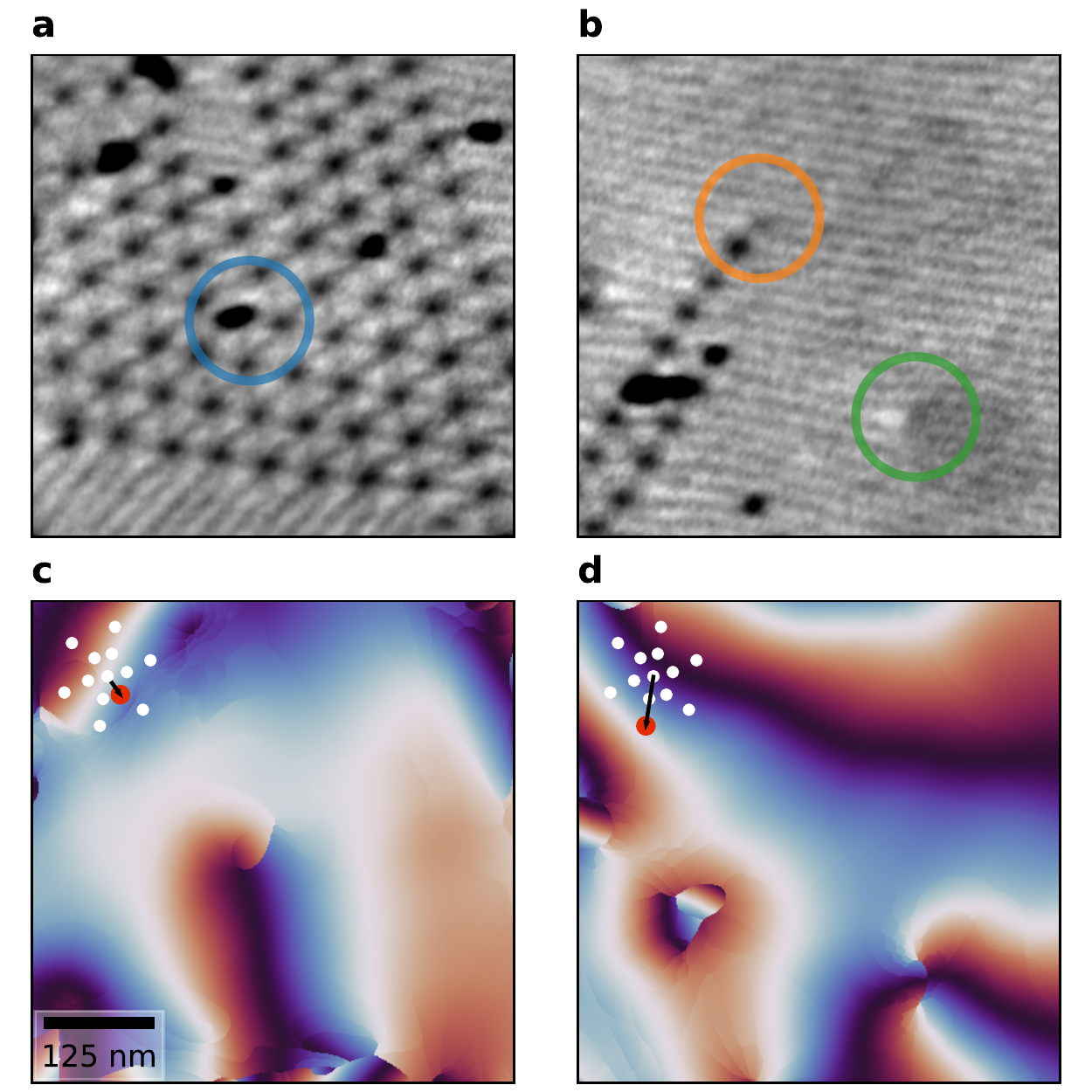}
\caption{\textbf{Dislocations in Sample B}
\subf{a} Dislocation in a triangular domain area at the center of the blue circle.
\subf{b} Two dislocations in a stripe domain area, indicated by orange and green circles.
\subf{c} GPA phases of \subref{a}, where the dislocation corresponds to a singularity in the GPA phase.
\subf{d} Similarly, GPA phase for the stripe domain in \subref{b}. The singularity of the dislocation in the lower right is of order 2, as the Burgers vector is parallel to the lattice vector corresponding to the GPA phase. 
The singularity at the end of the AA-node chain in the upper left is not visible in the GPA phase, as the Burgers vector is perpendicular to the to the relevant lattice vector.
Insets of (c,d) indicate the corresponding reciprocal vectors of the GPA phases. scalebar in \subref{c} applies to all panels.}\label{fig:GPAGSiCdisl}
\end{figure}

Edge dislocations in highly ordered areas in sample B, both in triangular and stripe areas show minimal distortion of the surrounding lattice.
Using GPA, we can highlight these atomic edge dislocations, just like in TBG, as shown in \figref{GPAGSiCdisl}, except in the case when the Burgers vector corresponding to the dislocation is parallel to the stripes.

As the density of dislocations is not dominated by edge dislocations in the buffer-layer / SiC substrate itself, this should be another indicator of the quality of the graphene layer.
In particular in sample C, high numbers of dislocations within a single stacking domain are found between areas of different stripe directions. 
This suggests that stripe domains might already form during growth with different stripe directions and therefore both different strain directions and a lattice mismatch. As they continue to grow and coalesce, this lattice mismatch can only be reconciled with edge dislocations, reminiscent of the rotational domain boundaries in CVD graphene, but with a much smaller lattice mismatch and therefore much lower number of edge dislocations.

\section{Conclusion and outlook}

In this work, we have shown that LEEM imaging of domain boundaries in epitaxial graphene on SiC enables the study of strain and atomic edge dislocations on large scales.
We found that the growth conditions of high quality graphene on SiC cause areas of anisotropic stripes in different directions. We have shown that these stripe domains might already form during the nucleation phase of the growth (as opposed to during cooldown) and cause atomic edge dislocations when different directions grow together. 

The growth temperature, growth duration and the amount of carbon pre-deposited all have significant effects on the growth, but also on the domain boundaries formed. Thus the study of these domain boundaries can aid the optimization of growth parameters.
In addition to these known parameters influencing the growth, we have seen strong indications that the direction of the step edges of the substrate with respect to the atomic lattice influences the stacking domains and therefore this miscut direction also influences the quality of the resulting graphene.
Finally, similar to the TBG case, we suspect that the topological defects in the domain boundary network could be interesting in itself, e.g. for their local electronic properties.

In this work, we have only scratched the surface of the information available in these domain boundary datasets. Therefore we here give a few more suggestions of information that could be extracted, but are beyond the scope of this work.
First, it would be informative to connect the images directly to the atomic lattice directions, either by connecting to LEED data, or possibly by observing the local directions of the substrate step edges.
Second, for the triangular domains the local uniaxial strain, biaxial strain, and twist should be separated using a Kerelsky-style decomposition based on the extracted $k$-vectors as described for the TBG case in Ref.~\cite{Kerelsky2019,de_jong_imaging_2021}.
Third, statistics of left versus right orientation spiral domain walls as a function of the local (minimal) twist angle between the lattices could be obtained. These could be used to measure energy differences as a function of twist angle and strain.
Finally, it would be worthwhile to use AC-LEEM to observe dynamics of domain walls, to study the stability of the orientation of spiral domain walls in the twist-free case and potentially obtain more detailed experimental data on the energy landscape that governs these domain boundaries.

\section*{Acknowledgements}
We thank Marcel Hesselberth and Douwe Scholma for their indispensable technical support.
We thank  Christian Ott, Heiko Weber, Robbert Schoo, Cees Flipse, Anna Sinterhauf and Martin Wenderoth for sharing their samples. This work was supported by the Netherlands Organisation for Scientific Research (NWO/OCW) as part of the Frontiers of Nanoscience program.
\bibliography{morphology.bib}

\begin{thebibliography}{46}%
\makeatletter
\providecommand \@ifxundefined [1]{%
 \@ifx{#1\undefined}
}%
\providecommand \@ifnum [1]{%
 \ifnum #1\expandafter \@firstoftwo
 \else \expandafter \@secondoftwo
 \fi
}%
\providecommand \@ifx [1]{%
 \ifx #1\expandafter \@firstoftwo
 \else \expandafter \@secondoftwo
 \fi
}%
\providecommand \natexlab [1]{#1}%
\providecommand \enquote  [1]{``#1''}%
\providecommand \bibnamefont  [1]{#1}%
\providecommand \bibfnamefont [1]{#1}%
\providecommand \citenamefont [1]{#1}%
\providecommand \href@noop [0]{\@secondoftwo}%
\providecommand \href [0]{\begingroup \@sanitize@url \@href}%
\providecommand \@href[1]{\@@startlink{#1}\@@href}%
\providecommand \@@href[1]{\endgroup#1\@@endlink}%
\providecommand \@sanitize@url [0]{\catcode `\\12\catcode `\$12\catcode
  `\&12\catcode `\#12\catcode `\^12\catcode `\_12\catcode `\%12\relax}%
\providecommand \@@startlink[1]{}%
\providecommand \@@endlink[0]{}%
\providecommand \url  [0]{\begingroup\@sanitize@url \@url }%
\providecommand \@url [1]{\endgroup\@href {#1}{\urlprefix }}%
\providecommand \urlprefix  [0]{URL }%
\providecommand \Eprint [0]{\href }%
\providecommand \doibase [0]{https://doi.org/}%
\providecommand \selectlanguage [0]{\@gobble}%
\providecommand \bibinfo  [0]{\@secondoftwo}%
\providecommand \bibfield  [0]{\@secondoftwo}%
\providecommand \translation [1]{[#1]}%
\providecommand \BibitemOpen [0]{}%
\providecommand \bibitemStop [0]{}%
\providecommand \bibitemNoStop [0]{.\EOS\space}%
\providecommand \EOS [0]{\spacefactor3000\relax}%
\providecommand \BibitemShut  [1]{\csname bibitem#1\endcsname}%
\let\auto@bib@innerbib\@empty
%</preamble>
\bibitem [{\citenamefont {Van~Bommel}\ \emph {et~al.}(1975)\citenamefont
  {Van~Bommel}, \citenamefont {Crombeen},\ and\ \citenamefont
  {Van~Tooren}}]{van_bommel_leed_1975}%
  \BibitemOpen
  \bibfield  {author} {\bibinfo {author} {\bibfnamefont {A.~J.}\ \bibnamefont
  {Van~Bommel}}, \bibinfo {author} {\bibfnamefont {J.~E.}\ \bibnamefont
  {Crombeen}},\ and\ \bibinfo {author} {\bibfnamefont {A.}~\bibnamefont
  {Van~Tooren}},\ }\bibfield  {title} {\bibinfo {title} {{LEED} and {Auger}
  electron observations of the {SiC}(0001) surface},\ }\href
  {https://doi.org/10.1016/0039-6028(75)90419-7} {\bibfield  {journal}
  {\bibinfo  {journal} {Surface Science}\ }\textbf {\bibinfo {volume} {48}},\
  \bibinfo {pages} {463} (\bibinfo {year} {1975})}\BibitemShut {NoStop}%
\bibitem [{\citenamefont {Riedl}\ \emph {et~al.}(2007)\citenamefont {Riedl},
  \citenamefont {Starke}, \citenamefont {Bernhardt}, \citenamefont {Franke},\
  and\ \citenamefont {Heinz}}]{riedl_structural_2007}%
  \BibitemOpen
  \bibfield  {author} {\bibinfo {author} {\bibfnamefont {C.}~\bibnamefont
  {Riedl}}, \bibinfo {author} {\bibfnamefont {U.}~\bibnamefont {Starke}},
  \bibinfo {author} {\bibfnamefont {J.}~\bibnamefont {Bernhardt}}, \bibinfo
  {author} {\bibfnamefont {M.}~\bibnamefont {Franke}},\ and\ \bibinfo {author}
  {\bibfnamefont {K.}~\bibnamefont {Heinz}},\ }\bibfield  {title} {\bibinfo
  {title} {Structural properties of the graphene-{SiC}(0001) interface as a key
  for the preparation of homogeneous large-terrace graphene surfaces},\ }\href
  {https://doi.org/10.1103/PhysRevB.76.245406} {\bibfield  {journal} {\bibinfo
  {journal} {Physical Review B}\ }\textbf {\bibinfo {volume} {76}},\ \bibinfo
  {pages} {245406} (\bibinfo {year} {2007})}\BibitemShut {NoStop}%
\bibitem [{\citenamefont {Tanaka}\ \emph {et~al.}(2010)\citenamefont {Tanaka},
  \citenamefont {Morita},\ and\ \citenamefont
  {Hibino}}]{tanaka_anisotropic_2010}%
  \BibitemOpen
  \bibfield  {author} {\bibinfo {author} {\bibfnamefont {S.}~\bibnamefont
  {Tanaka}}, \bibinfo {author} {\bibfnamefont {K.}~\bibnamefont {Morita}},\
  and\ \bibinfo {author} {\bibfnamefont {H.}~\bibnamefont {Hibino}},\
  }\bibfield  {title} {\bibinfo {title} {Anisotropic layer-by-layer growth of
  graphene on vicinal {SiC}(0001) surfaces},\ }\href
  {https://doi.org/10.1103/PhysRevB.81.041406} {\bibfield  {journal} {\bibinfo
  {journal} {Physical Review B}\ }\textbf {\bibinfo {volume} {81}},\ \bibinfo
  {pages} {041406} (\bibinfo {year} {2010})}\BibitemShut {NoStop}%
\bibitem [{\citenamefont {Tromp}\ and\ \citenamefont
  {Hannon}(2009)}]{tromp_thermodynamics_2009}%
  \BibitemOpen
  \bibfield  {author} {\bibinfo {author} {\bibfnamefont {R.~M.}\ \bibnamefont
  {Tromp}}\ and\ \bibinfo {author} {\bibfnamefont {J.~B.}\ \bibnamefont
  {Hannon}},\ }\bibfield  {title} {\bibinfo {title} {Thermodynamics and
  kinetics of graphene growth on {SiC}(0001)},\ }\href
  {https://doi.org/10.1103/PhysRevLett.102.106104} {\bibfield  {journal}
  {\bibinfo  {journal} {Physical Review Letters}\ }\textbf {\bibinfo {volume}
  {102}},\ \bibinfo {pages} {106104} (\bibinfo {year} {2009})}\BibitemShut
  {NoStop}%
\bibitem [{\citenamefont {Emtsev}\ \emph {et~al.}(2009)\citenamefont {Emtsev},
  \citenamefont {Bostwick}, \citenamefont {Horn}, \citenamefont {Jobst},
  \citenamefont {Kellogg}, \citenamefont {Ley}, \citenamefont {McChesney},
  \citenamefont {Ohta}, \citenamefont {Reshanov}, \citenamefont {Röhrl},
  \citenamefont {Rotenberg}, \citenamefont {Schmid}, \citenamefont {Waldmann},
  \citenamefont {Weber},\ and\ \citenamefont {Seyller}}]{emtsev2009-towards}%
  \BibitemOpen
  \bibfield  {author} {\bibinfo {author} {\bibfnamefont {K.~V.}\ \bibnamefont
  {Emtsev}}, \bibinfo {author} {\bibfnamefont {A.}~\bibnamefont {Bostwick}},
  \bibinfo {author} {\bibfnamefont {K.}~\bibnamefont {Horn}}, \bibinfo {author}
  {\bibfnamefont {J.}~\bibnamefont {Jobst}}, \bibinfo {author} {\bibfnamefont
  {G.~L.}\ \bibnamefont {Kellogg}}, \bibinfo {author} {\bibfnamefont
  {L.}~\bibnamefont {Ley}}, \bibinfo {author} {\bibfnamefont {J.~L.}\
  \bibnamefont {McChesney}}, \bibinfo {author} {\bibfnamefont {T.}~\bibnamefont
  {Ohta}}, \bibinfo {author} {\bibfnamefont {S.~A.}\ \bibnamefont {Reshanov}},
  \bibinfo {author} {\bibfnamefont {J.}~\bibnamefont {Röhrl}}, \bibinfo
  {author} {\bibfnamefont {E.}~\bibnamefont {Rotenberg}}, \bibinfo {author}
  {\bibfnamefont {A.~K.}\ \bibnamefont {Schmid}}, \bibinfo {author}
  {\bibfnamefont {D.}~\bibnamefont {Waldmann}}, \bibinfo {author}
  {\bibfnamefont {H.~B.}\ \bibnamefont {Weber}},\ and\ \bibinfo {author}
  {\bibfnamefont {T.}~\bibnamefont {Seyller}},\ }\bibfield  {title} {\bibinfo
  {title} {Towards wafer-size graphene layers by atmospheric pressure
  graphitization of silicon carbide},\ }\href
  {https://doi.org/10.1038/nmat2382} {\bibfield  {journal} {\bibinfo  {journal}
  {Nature Materials}\ }\textbf {\bibinfo {volume} {8}},\ \bibinfo {pages} {203}
  (\bibinfo {year} {2009})}\BibitemShut {NoStop}%
\bibitem [{\citenamefont {Kruskopf}\ \emph {et~al.}(2016)\citenamefont
  {Kruskopf}, \citenamefont {Pakdehi}, \citenamefont {Pierz}, \citenamefont
  {Wundrack}, \citenamefont {Stosch}, \citenamefont {Dziomba}, \citenamefont
  {Götz}, \citenamefont {Baringhaus}, \citenamefont {Aprojanz}, \citenamefont
  {Tegenkamp}, \citenamefont {Lidzba}, \citenamefont {Seyller}, \citenamefont
  {Hohls}, \citenamefont {Ahlers},\ and\ \citenamefont
  {Schumacher}}]{kruskopf_comeback_2016}%
  \BibitemOpen
  \bibfield  {author} {\bibinfo {author} {\bibfnamefont {M.}~\bibnamefont
  {Kruskopf}}, \bibinfo {author} {\bibfnamefont {D.~M.}\ \bibnamefont
  {Pakdehi}}, \bibinfo {author} {\bibfnamefont {K.}~\bibnamefont {Pierz}},
  \bibinfo {author} {\bibfnamefont {S.}~\bibnamefont {Wundrack}}, \bibinfo
  {author} {\bibfnamefont {R.}~\bibnamefont {Stosch}}, \bibinfo {author}
  {\bibfnamefont {T.}~\bibnamefont {Dziomba}}, \bibinfo {author} {\bibfnamefont
  {M.}~\bibnamefont {Götz}}, \bibinfo {author} {\bibfnamefont
  {J.}~\bibnamefont {Baringhaus}}, \bibinfo {author} {\bibfnamefont
  {J.}~\bibnamefont {Aprojanz}}, \bibinfo {author} {\bibfnamefont
  {C.}~\bibnamefont {Tegenkamp}}, \bibinfo {author} {\bibfnamefont
  {J.}~\bibnamefont {Lidzba}}, \bibinfo {author} {\bibfnamefont
  {T.}~\bibnamefont {Seyller}}, \bibinfo {author} {\bibfnamefont
  {F.}~\bibnamefont {Hohls}}, \bibinfo {author} {\bibfnamefont {F.~J.}\
  \bibnamefont {Ahlers}},\ and\ \bibinfo {author} {\bibfnamefont {H.~W.}\
  \bibnamefont {Schumacher}},\ }\bibfield  {title} {\bibinfo {title} {Comeback
  of epitaxial graphene for electronics: large-area growth of bilayer-free
  graphene on {SiC}},\ }\href {https://doi.org/10.1088/2053-1583/3/4/041002}
  {\bibfield  {journal} {\bibinfo  {journal} {2D Materials}\ }\textbf {\bibinfo
  {volume} {3}},\ \bibinfo {pages} {041002} (\bibinfo {year}
  {2016})}\BibitemShut {NoStop}%
\bibitem [{\citenamefont {Momeni~Pakdehi}\ \emph {et~al.}(2018)\citenamefont
  {Momeni~Pakdehi}, \citenamefont {Aprojanz}, \citenamefont {Sinterhauf},
  \citenamefont {Pierz}, \citenamefont {Kruskopf}, \citenamefont {Willke},
  \citenamefont {Baringhaus}, \citenamefont {Stöckmann}, \citenamefont
  {Traeger}, \citenamefont {Hohls}, \citenamefont {Tegenkamp}, \citenamefont
  {Wenderoth}, \citenamefont {Ahlers},\ and\ \citenamefont
  {Schumacher}}]{momeni_pakdehi_minimum_2018}%
  \BibitemOpen
  \bibfield  {author} {\bibinfo {author} {\bibfnamefont {D.}~\bibnamefont
  {Momeni~Pakdehi}}, \bibinfo {author} {\bibfnamefont {J.}~\bibnamefont
  {Aprojanz}}, \bibinfo {author} {\bibfnamefont {A.}~\bibnamefont
  {Sinterhauf}}, \bibinfo {author} {\bibfnamefont {K.}~\bibnamefont {Pierz}},
  \bibinfo {author} {\bibfnamefont {M.}~\bibnamefont {Kruskopf}}, \bibinfo
  {author} {\bibfnamefont {P.}~\bibnamefont {Willke}}, \bibinfo {author}
  {\bibfnamefont {J.}~\bibnamefont {Baringhaus}}, \bibinfo {author}
  {\bibfnamefont {J.~P.}\ \bibnamefont {Stöckmann}}, \bibinfo {author}
  {\bibfnamefont {G.~A.}\ \bibnamefont {Traeger}}, \bibinfo {author}
  {\bibfnamefont {F.}~\bibnamefont {Hohls}}, \bibinfo {author} {\bibfnamefont
  {C.}~\bibnamefont {Tegenkamp}}, \bibinfo {author} {\bibfnamefont
  {M.}~\bibnamefont {Wenderoth}}, \bibinfo {author} {\bibfnamefont {F.~J.}\
  \bibnamefont {Ahlers}},\ and\ \bibinfo {author} {\bibfnamefont {H.~W.}\
  \bibnamefont {Schumacher}},\ }\bibfield  {title} {\bibinfo {title} {Minimum
  {Resistance} {Anisotropy} of {Epitaxial} {Graphene} on {SiC}},\ }\href
  {https://doi.org/10.1021/acsami.7b18641} {\bibfield  {journal} {\bibinfo
  {journal} {ACS Applied Materials \& Interfaces}\ }\textbf {\bibinfo {volume}
  {10}},\ \bibinfo {pages} {6039} (\bibinfo {year} {2018})}\BibitemShut
  {NoStop}%
\bibitem [{\citenamefont {Sonde}\ \emph {et~al.}(2010)\citenamefont {Sonde},
  \citenamefont {Giannazzo}, \citenamefont {Vecchio}, \citenamefont {Yakimova},
  \citenamefont {Rimini},\ and\ \citenamefont {Raineri}}]{sonde_role_2010}%
  \BibitemOpen
  \bibfield  {author} {\bibinfo {author} {\bibfnamefont {S.}~\bibnamefont
  {Sonde}}, \bibinfo {author} {\bibfnamefont {F.}~\bibnamefont {Giannazzo}},
  \bibinfo {author} {\bibfnamefont {C.}~\bibnamefont {Vecchio}}, \bibinfo
  {author} {\bibfnamefont {R.}~\bibnamefont {Yakimova}}, \bibinfo {author}
  {\bibfnamefont {E.}~\bibnamefont {Rimini}},\ and\ \bibinfo {author}
  {\bibfnamefont {V.}~\bibnamefont {Raineri}},\ }\bibfield  {title} {\bibinfo
  {title} {Role of graphene/substrate interface on the local transport
  properties of the two-dimensional electron gas},\ }\href
  {https://doi.org/10.1063/1.3489942} {\bibfield  {journal} {\bibinfo
  {journal} {Applied Physics Letters}\ }\textbf {\bibinfo {volume} {97}},\
  \bibinfo {pages} {132101} (\bibinfo {year} {2010})}\BibitemShut {NoStop}%
\bibitem [{\citenamefont {Kim}\ \emph {et~al.}(2008)\citenamefont {Kim},
  \citenamefont {Ihm}, \citenamefont {Choi},\ and\ \citenamefont
  {Son}}]{kim_origin_2008}%
  \BibitemOpen
  \bibfield  {author} {\bibinfo {author} {\bibfnamefont {S.}~\bibnamefont
  {Kim}}, \bibinfo {author} {\bibfnamefont {J.}~\bibnamefont {Ihm}}, \bibinfo
  {author} {\bibfnamefont {H.~J.}\ \bibnamefont {Choi}},\ and\ \bibinfo
  {author} {\bibfnamefont {Y.-W.}\ \bibnamefont {Son}},\ }\bibfield  {title}
  {\bibinfo {title} {Origin of {Anomalous} {Electronic} {Structures} of
  {Epitaxial} {Graphene} on {Silicon} {Carbide}},\ }\href
  {https://doi.org/10.1103/PhysRevLett.100.176802} {\bibfield  {journal}
  {\bibinfo  {journal} {Physical Review Letters}\ }\textbf {\bibinfo {volume}
  {100}},\ \bibinfo {pages} {176802} (\bibinfo {year} {2008})}\BibitemShut
  {NoStop}%
\bibitem [{\citenamefont {de~Jong}\ \emph {et~al.}(2018)\citenamefont
  {de~Jong}, \citenamefont {Krasovskii}, \citenamefont {Ott}, \citenamefont
  {Tromp}, \citenamefont {van~der Molen},\ and\ \citenamefont
  {Jobst}}]{dejong2018intrinsic}%
  \BibitemOpen
  \bibfield  {author} {\bibinfo {author} {\bibfnamefont {T.~A.}\ \bibnamefont
  {de~Jong}}, \bibinfo {author} {\bibfnamefont {E.~E.}\ \bibnamefont
  {Krasovskii}}, \bibinfo {author} {\bibfnamefont {C.}~\bibnamefont {Ott}},
  \bibinfo {author} {\bibfnamefont {R.~M.}\ \bibnamefont {Tromp}}, \bibinfo
  {author} {\bibfnamefont {S.~J.}\ \bibnamefont {van~der Molen}},\ and\
  \bibinfo {author} {\bibfnamefont {J.}~\bibnamefont {Jobst}},\ }\bibfield
  {title} {\bibinfo {title} {Intrinsic stacking domains in graphene on silicon
  carbide: {A} pathway for intercalation},\ }\href
  {https://doi.org/10.1103/PhysRevMaterials.2.104005} {\bibfield  {journal}
  {\bibinfo  {journal} {Physical Review Materials}\ }\textbf {\bibinfo {volume}
  {2}},\ \bibinfo {pages} {104005} (\bibinfo {year} {2018})}\BibitemShut
  {NoStop}%
\bibitem [{\citenamefont {Cho}\ \emph {et~al.}(2013)\citenamefont {Cho},
  \citenamefont {Kang}, \citenamefont {Kim}, \citenamefont {Lee}, \citenamefont
  {Woo}, \citenamefont {Kong}, \citenamefont {Kim}, \citenamefont {Kim},
  \citenamefont {Zhang}, \citenamefont {Stroscio}, \citenamefont {Kim},\ and\
  \citenamefont {Lyeo}}]{cho_thermoelectric_2013}%
  \BibitemOpen
  \bibfield  {author} {\bibinfo {author} {\bibfnamefont {S.}~\bibnamefont
  {Cho}}, \bibinfo {author} {\bibfnamefont {S.~D.}\ \bibnamefont {Kang}},
  \bibinfo {author} {\bibfnamefont {W.}~\bibnamefont {Kim}}, \bibinfo {author}
  {\bibfnamefont {E.-S.}\ \bibnamefont {Lee}}, \bibinfo {author} {\bibfnamefont
  {S.-J.}\ \bibnamefont {Woo}}, \bibinfo {author} {\bibfnamefont {K.-J.}\
  \bibnamefont {Kong}}, \bibinfo {author} {\bibfnamefont {I.}~\bibnamefont
  {Kim}}, \bibinfo {author} {\bibfnamefont {H.-D.}\ \bibnamefont {Kim}},
  \bibinfo {author} {\bibfnamefont {T.}~\bibnamefont {Zhang}}, \bibinfo
  {author} {\bibfnamefont {J.~A.}\ \bibnamefont {Stroscio}}, \bibinfo {author}
  {\bibfnamefont {Y.-H.}\ \bibnamefont {Kim}},\ and\ \bibinfo {author}
  {\bibfnamefont {H.-K.}\ \bibnamefont {Lyeo}},\ }\bibfield  {title} {\bibinfo
  {title} {Thermoelectric imaging of structural disorder in epitaxial
  graphene},\ }\href {https://doi.org/10.1038/nmat3708} {\bibfield  {journal}
  {\bibinfo  {journal} {Nature Materials}\ }\textbf {\bibinfo {volume} {12}},\
  \bibinfo {pages} {913} (\bibinfo {year} {2013})}\BibitemShut {NoStop}%
\bibitem [{\citenamefont {Hibino}\ \emph
  {et~al.}(2009{\natexlab{a}})\citenamefont {Hibino}, \citenamefont {Mizuno},
  \citenamefont {Kageshima}, \citenamefont {Nagase},\ and\ \citenamefont
  {Yamaguchi}}]{hibino_stacking_2009}%
  \BibitemOpen
  \bibfield  {author} {\bibinfo {author} {\bibfnamefont {H.}~\bibnamefont
  {Hibino}}, \bibinfo {author} {\bibfnamefont {S.}~\bibnamefont {Mizuno}},
  \bibinfo {author} {\bibfnamefont {H.}~\bibnamefont {Kageshima}}, \bibinfo
  {author} {\bibfnamefont {M.}~\bibnamefont {Nagase}},\ and\ \bibinfo {author}
  {\bibfnamefont {H.}~\bibnamefont {Yamaguchi}},\ }\bibfield  {title} {\bibinfo
  {title} {Stacking domains of epitaxial few-layer graphene on {SiC}(0001)},\
  }\href {https://doi.org/10.1103/PhysRevB.80.085406} {\bibfield  {journal}
  {\bibinfo  {journal} {Physical Review B}\ }\textbf {\bibinfo {volume} {80}},\
  \bibinfo {pages} {085406} (\bibinfo {year} {2009}{\natexlab{a}})}\BibitemShut
  {NoStop}%
\bibitem [{\citenamefont {Lalmi}\ \emph {et~al.}(2014)\citenamefont {Lalmi},
  \citenamefont {Girard}, \citenamefont {Pallecchi}, \citenamefont {Silly},
  \citenamefont {David}, \citenamefont {Latil}, \citenamefont {Sirotti},\ and\
  \citenamefont {Ouerghi}}]{lalmi_flower-shaped_2014}%
  \BibitemOpen
  \bibfield  {author} {\bibinfo {author} {\bibfnamefont {B.}~\bibnamefont
  {Lalmi}}, \bibinfo {author} {\bibfnamefont {J.~C.}\ \bibnamefont {Girard}},
  \bibinfo {author} {\bibfnamefont {E.}~\bibnamefont {Pallecchi}}, \bibinfo
  {author} {\bibfnamefont {M.}~\bibnamefont {Silly}}, \bibinfo {author}
  {\bibfnamefont {C.}~\bibnamefont {David}}, \bibinfo {author} {\bibfnamefont
  {S.}~\bibnamefont {Latil}}, \bibinfo {author} {\bibfnamefont
  {F.}~\bibnamefont {Sirotti}},\ and\ \bibinfo {author} {\bibfnamefont
  {A.}~\bibnamefont {Ouerghi}},\ }\bibfield  {title} {\bibinfo {title}
  {Flower-{Shaped} {Domains} and {Wrinkles} in {Trilayer} {Epitaxial}
  {Graphene} on {Silicon} {Carbide}},\ }\href
  {https://doi.org/10.1038/srep04066} {\bibfield  {journal} {\bibinfo
  {journal} {Scientific Reports}\ }\textbf {\bibinfo {volume} {4}},\ \bibinfo
  {pages} {4066} (\bibinfo {year} {2014})}\BibitemShut {NoStop}%
\bibitem [{\citenamefont {de~Jong}\ \emph
  {et~al.}(2022{\natexlab{a}})\citenamefont {de~Jong}, \citenamefont {Chen},
  \citenamefont {Krasovskii}, \citenamefont {Tromp}, \citenamefont {Jobst},\
  and\ \citenamefont {van~der Molen}}]{de_jong_2022_contrast}%
  \BibitemOpen
  \bibfield  {author} {\bibinfo {author} {\bibfnamefont {T.~A.}\ \bibnamefont
  {de~Jong}}, \bibinfo {author} {\bibfnamefont {X.}~\bibnamefont {Chen}},
  \bibinfo {author} {\bibfnamefont {E.~E.}\ \bibnamefont {Krasovskii}},
  \bibinfo {author} {\bibfnamefont {R.~M.}\ \bibnamefont {Tromp}}, \bibinfo
  {author} {\bibfnamefont {J.}~\bibnamefont {Jobst}},\ and\ \bibinfo {author}
  {\bibfnamefont {S.~J.}\ \bibnamefont {van~der Molen}},\ }\bibfield  {title}
  {\bibinfo {title} {Low-energy electron microscopy contrast of stacking
  boundaries: comparing twisted few-layer graphene and strained epitaxial
  graphene on silicon carbide.},\ }\href@noop {} {\bibfield  {journal}
  {\bibinfo  {journal} {arXiv [cond-mat]}\ } (\bibinfo {year}
  {2022}{\natexlab{a}})}\BibitemShut {NoStop}%
\bibitem [{\citenamefont {de~Jong}\ \emph
  {et~al.}(2022{\natexlab{b}})\citenamefont {de~Jong}, \citenamefont
  {Benschop}, \citenamefont {Chen}, \citenamefont {Krasovskii}, \citenamefont
  {de~Dood}, \citenamefont {Tromp}, \citenamefont {Allan},\ and\ \citenamefont
  {van~der Molen}}]{de_jong_imaging_2021}%
  \BibitemOpen
  \bibfield  {author} {\bibinfo {author} {\bibfnamefont {T.~A.}\ \bibnamefont
  {de~Jong}}, \bibinfo {author} {\bibfnamefont {T.}~\bibnamefont {Benschop}},
  \bibinfo {author} {\bibfnamefont {X.}~\bibnamefont {Chen}}, \bibinfo {author}
  {\bibfnamefont {E.~E.}\ \bibnamefont {Krasovskii}}, \bibinfo {author}
  {\bibfnamefont {M.~J.~A.}\ \bibnamefont {de~Dood}}, \bibinfo {author}
  {\bibfnamefont {R.~M.}\ \bibnamefont {Tromp}}, \bibinfo {author}
  {\bibfnamefont {M.~P.}\ \bibnamefont {Allan}},\ and\ \bibinfo {author}
  {\bibfnamefont {S.~J.}\ \bibnamefont {van~der Molen}},\ }\bibfield  {title}
  {\bibinfo {title} {Imaging moiré deformation and dynamics in twisted bilayer
  graphene},\ }\bibfield  {journal} {\bibinfo  {journal} {Nature
  Communications}\ }\textbf {\bibinfo {volume} {13}},\ \href
  {https://doi.org/10.1038/s41467-021-27646-1} {10.1038/s41467-021-27646-1}
  (\bibinfo {year} {2022}{\natexlab{b}})\BibitemShut {NoStop}%
\bibitem [{\citenamefont {Ott}(2021)}]{ott_light_2021}%
  \BibitemOpen
  \bibfield  {author} {\bibinfo {author} {\bibfnamefont {C.}~\bibnamefont
  {Ott}},\ }\href
  {https://opus4.kobv.de/opus4-fau/files/16722/ChristianOttDissertation.pdf}
  {\bibinfo {title} {Light and electron emission at epitaxial graphene
  nano-contacts}} (\bibinfo {year} {2021})\BibitemShut {NoStop}%
\bibitem [{\citenamefont {Yazdi}\ \emph {et~al.}(2013)\citenamefont {Yazdi},
  \citenamefont {Vasiliauskas}, \citenamefont {Iakimov}, \citenamefont
  {Zakharov}, \citenamefont {Syväjärvi},\ and\ \citenamefont
  {Yakimova}}]{yazdi_growth_2013}%
  \BibitemOpen
  \bibfield  {author} {\bibinfo {author} {\bibfnamefont {G.~R.}\ \bibnamefont
  {Yazdi}}, \bibinfo {author} {\bibfnamefont {R.}~\bibnamefont {Vasiliauskas}},
  \bibinfo {author} {\bibfnamefont {T.}~\bibnamefont {Iakimov}}, \bibinfo
  {author} {\bibfnamefont {A.}~\bibnamefont {Zakharov}}, \bibinfo {author}
  {\bibfnamefont {M.}~\bibnamefont {Syväjärvi}},\ and\ \bibinfo {author}
  {\bibfnamefont {R.}~\bibnamefont {Yakimova}},\ }\bibfield  {title} {\bibinfo
  {title} {Growth of large area monolayer graphene on {3C}-{SiC} and a
  comparison with other {SiC} polytypes},\ }\href
  {https://doi.org/10.1016/j.carbon.2013.02.022} {\bibfield  {journal}
  {\bibinfo  {journal} {Carbon}\ }\textbf {\bibinfo {volume} {57}},\ \bibinfo
  {pages} {477} (\bibinfo {year} {2013})}\BibitemShut {NoStop}%
\bibitem [{\citenamefont {de~Jong}(2022)}]{graphene-stacking-domains-code}%
  \BibitemOpen
  \bibfield  {author} {\bibinfo {author} {\bibfnamefont {T.~A.}\ \bibnamefont
  {de~Jong}},\ }\href
  {https://github.com/TAdeJong/graphene-stacking-domains-code} {\bibinfo
  {title} {Graphene stacking domains code}} (\bibinfo {year}
  {2022})\BibitemShut {NoStop}%
\bibitem [{\citenamefont {Momeni~Pakdehi}\ \emph {et~al.}()\citenamefont
  {Momeni~Pakdehi}, \citenamefont {Schädlich}, \citenamefont {Nguyen},
  \citenamefont {Zakharov}, \citenamefont {Wundrack}, \citenamefont
  {Najafidehaghani}, \citenamefont {Speck}, \citenamefont {Pierz},
  \citenamefont {Seyller}, \citenamefont {Tegenkamp},\ and\ \citenamefont
  {Schumacher}}]{momeni_pakdehi_silicon_2020}%
  \BibitemOpen
  \bibfield  {author} {\bibinfo {author} {\bibfnamefont {D.}~\bibnamefont
  {Momeni~Pakdehi}}, \bibinfo {author} {\bibfnamefont {P.}~\bibnamefont
  {Schädlich}}, \bibinfo {author} {\bibfnamefont {T.~T.~N.}\ \bibnamefont
  {Nguyen}}, \bibinfo {author} {\bibfnamefont {A.~A.}\ \bibnamefont
  {Zakharov}}, \bibinfo {author} {\bibfnamefont {S.}~\bibnamefont {Wundrack}},
  \bibinfo {author} {\bibfnamefont {E.}~\bibnamefont {Najafidehaghani}},
  \bibinfo {author} {\bibfnamefont {F.}~\bibnamefont {Speck}}, \bibinfo
  {author} {\bibfnamefont {K.}~\bibnamefont {Pierz}}, \bibinfo {author}
  {\bibfnamefont {T.}~\bibnamefont {Seyller}}, \bibinfo {author} {\bibfnamefont
  {C.}~\bibnamefont {Tegenkamp}},\ and\ \bibinfo {author} {\bibfnamefont
  {H.~W.}\ \bibnamefont {Schumacher}},\ }\bibfield  {title} {\bibinfo {title}
  {Silicon {Carbide} {Stacking}-{Order}-{Induced} {Doping} {Variation} in
  {Epitaxial} {Graphene}},\ }\href {https://doi.org/10.1002/adfm.202004695}
  {\bibfield  {journal} {\bibinfo  {journal} {Advanced Functional Materials}\
  }\textbf {\bibinfo {volume} {30}},\ \bibinfo {pages} {2004695}}\BibitemShut
  {NoStop}%
\bibitem [{\citenamefont {Sinterhauf}\ \emph {et~al.}(2020)\citenamefont
  {Sinterhauf}, \citenamefont {Traeger}, \citenamefont {Momeni~Pakdehi},
  \citenamefont {Schädlich}, \citenamefont {Willke}, \citenamefont {Speck},
  \citenamefont {Seyller}, \citenamefont {Tegenkamp}, \citenamefont {Pierz},
  \citenamefont {Schumacher},\ and\ \citenamefont
  {Wenderoth}}]{sinterhauf_substrate_2020}%
  \BibitemOpen
  \bibfield  {author} {\bibinfo {author} {\bibfnamefont {A.}~\bibnamefont
  {Sinterhauf}}, \bibinfo {author} {\bibfnamefont {G.~A.}\ \bibnamefont
  {Traeger}}, \bibinfo {author} {\bibfnamefont {D.}~\bibnamefont
  {Momeni~Pakdehi}}, \bibinfo {author} {\bibfnamefont {P.}~\bibnamefont
  {Schädlich}}, \bibinfo {author} {\bibfnamefont {P.}~\bibnamefont {Willke}},
  \bibinfo {author} {\bibfnamefont {F.}~\bibnamefont {Speck}}, \bibinfo
  {author} {\bibfnamefont {T.}~\bibnamefont {Seyller}}, \bibinfo {author}
  {\bibfnamefont {C.}~\bibnamefont {Tegenkamp}}, \bibinfo {author}
  {\bibfnamefont {K.}~\bibnamefont {Pierz}}, \bibinfo {author} {\bibfnamefont
  {H.~W.}\ \bibnamefont {Schumacher}},\ and\ \bibinfo {author} {\bibfnamefont
  {M.}~\bibnamefont {Wenderoth}},\ }\bibfield  {title} {\bibinfo {title}
  {Substrate induced nanoscale resistance variation in epitaxial graphene},\
  }\href {https://doi.org/10.1038/s41467-019-14192-0} {\bibfield  {journal}
  {\bibinfo  {journal} {Nature Communications}\ }\textbf {\bibinfo {volume}
  {11}},\ \bibinfo {pages} {555} (\bibinfo {year} {2020})}\BibitemShut
  {NoStop}%
\bibitem [{\citenamefont {Speck}\ \emph {et~al.}()\citenamefont {Speck},
  \citenamefont {Ostler}, \citenamefont {Besendörfer}, \citenamefont {Krone},
  \citenamefont {Wanke},\ and\ \citenamefont {Seyller}}]{speck2017growth}%
  \BibitemOpen
  \bibfield  {author} {\bibinfo {author} {\bibfnamefont {F.}~\bibnamefont
  {Speck}}, \bibinfo {author} {\bibfnamefont {M.}~\bibnamefont {Ostler}},
  \bibinfo {author} {\bibfnamefont {S.}~\bibnamefont {Besendörfer}}, \bibinfo
  {author} {\bibfnamefont {J.}~\bibnamefont {Krone}}, \bibinfo {author}
  {\bibfnamefont {M.}~\bibnamefont {Wanke}},\ and\ \bibinfo {author}
  {\bibfnamefont {T.}~\bibnamefont {Seyller}},\ }\bibfield  {title} {\bibinfo
  {title} {{Growth and Intercalation of Graphene on Silicon Carbide Studied by
  Low-Energy Electron Microscopy}},\ }\href
  {https://doi.org/10.1002/andp.201700046} {\bibfield  {journal} {\bibinfo
  {journal} {Annalen der Physik}\ }\textbf {\bibinfo {volume} {529}},\ \bibinfo
  {pages} {1700046}}\BibitemShut {NoStop}%
\bibitem [{\citenamefont {Butz}\ \emph {et~al.}(2014)\citenamefont {Butz},
  \citenamefont {Dolle}, \citenamefont {Niekiel}, \citenamefont {Weber},
  \citenamefont {Waldmann}, \citenamefont {Weber}, \citenamefont {Meyer},\ and\
  \citenamefont {Spiecker}}]{butz2014dislocations}%
  \BibitemOpen
  \bibfield  {author} {\bibinfo {author} {\bibfnamefont {B.}~\bibnamefont
  {Butz}}, \bibinfo {author} {\bibfnamefont {C.}~\bibnamefont {Dolle}},
  \bibinfo {author} {\bibfnamefont {F.}~\bibnamefont {Niekiel}}, \bibinfo
  {author} {\bibfnamefont {K.}~\bibnamefont {Weber}}, \bibinfo {author}
  {\bibfnamefont {D.}~\bibnamefont {Waldmann}}, \bibinfo {author}
  {\bibfnamefont {H.~B.}\ \bibnamefont {Weber}}, \bibinfo {author}
  {\bibfnamefont {B.}~\bibnamefont {Meyer}},\ and\ \bibinfo {author}
  {\bibfnamefont {E.}~\bibnamefont {Spiecker}},\ }\bibfield  {title} {\bibinfo
  {title} {{Dislocations in bilayer graphene}},\ }\href
  {https://doi.org/10.1038/nature12780} {\bibfield  {journal} {\bibinfo
  {journal} {Nature}\ }\textbf {\bibinfo {volume} {505}},\ \bibinfo {pages}
  {533} (\bibinfo {year} {2014})}\BibitemShut {NoStop}%
\bibitem [{\citenamefont {Halbertal}\ \emph {et~al.}(2021)\citenamefont
  {Halbertal}, \citenamefont {Finney}, \citenamefont {Sunku}, \citenamefont
  {Kerelsky}, \citenamefont {Rubio-Verdú}, \citenamefont {Shabani},
  \citenamefont {Xian}, \citenamefont {Carr}, \citenamefont {Chen},
  \citenamefont {Zhang}, \citenamefont {Wang}, \citenamefont
  {Gonzalez-Acevedo}, \citenamefont {McLeod}, \citenamefont {Rhodes},
  \citenamefont {Watanabe}, \citenamefont {Taniguchi}, \citenamefont {Kaxiras},
  \citenamefont {Dean}, \citenamefont {Hone}, \citenamefont {Pasupathy},
  \citenamefont {Kennes}, \citenamefont {Rubio},\ and\ \citenamefont
  {Basov}}]{halbertal_moire_2021}%
  \BibitemOpen
  \bibfield  {author} {\bibinfo {author} {\bibfnamefont {D.}~\bibnamefont
  {Halbertal}}, \bibinfo {author} {\bibfnamefont {N.~R.}\ \bibnamefont
  {Finney}}, \bibinfo {author} {\bibfnamefont {S.~S.}\ \bibnamefont {Sunku}},
  \bibinfo {author} {\bibfnamefont {A.}~\bibnamefont {Kerelsky}}, \bibinfo
  {author} {\bibfnamefont {C.}~\bibnamefont {Rubio-Verdú}}, \bibinfo {author}
  {\bibfnamefont {S.}~\bibnamefont {Shabani}}, \bibinfo {author} {\bibfnamefont
  {L.}~\bibnamefont {Xian}}, \bibinfo {author} {\bibfnamefont {S.}~\bibnamefont
  {Carr}}, \bibinfo {author} {\bibfnamefont {S.}~\bibnamefont {Chen}}, \bibinfo
  {author} {\bibfnamefont {C.}~\bibnamefont {Zhang}}, \bibinfo {author}
  {\bibfnamefont {L.}~\bibnamefont {Wang}}, \bibinfo {author} {\bibfnamefont
  {D.}~\bibnamefont {Gonzalez-Acevedo}}, \bibinfo {author} {\bibfnamefont
  {A.~S.}\ \bibnamefont {McLeod}}, \bibinfo {author} {\bibfnamefont
  {D.}~\bibnamefont {Rhodes}}, \bibinfo {author} {\bibfnamefont
  {K.}~\bibnamefont {Watanabe}}, \bibinfo {author} {\bibfnamefont
  {T.}~\bibnamefont {Taniguchi}}, \bibinfo {author} {\bibfnamefont
  {E.}~\bibnamefont {Kaxiras}}, \bibinfo {author} {\bibfnamefont {C.~R.}\
  \bibnamefont {Dean}}, \bibinfo {author} {\bibfnamefont {J.~C.}\ \bibnamefont
  {Hone}}, \bibinfo {author} {\bibfnamefont {A.~N.}\ \bibnamefont {Pasupathy}},
  \bibinfo {author} {\bibfnamefont {D.~M.}\ \bibnamefont {Kennes}}, \bibinfo
  {author} {\bibfnamefont {A.}~\bibnamefont {Rubio}},\ and\ \bibinfo {author}
  {\bibfnamefont {D.~N.}\ \bibnamefont {Basov}},\ }\bibfield  {title} {\bibinfo
  {title} {Moiré metrology of energy landscapes in van der {Waals}
  heterostructures},\ }\href {https://doi.org/10.1038/s41467-020-20428-1}
  {\bibfield  {journal} {\bibinfo  {journal} {Nature Communications}\ }\textbf
  {\bibinfo {volume} {12}},\ \bibinfo {pages} {242} (\bibinfo {year}
  {2021})}\BibitemShut {NoStop}%
\bibitem [{\citenamefont {Guerrero-Avilés}\ \emph {et~al.}(2021)\citenamefont
  {Guerrero-Avilés}, \citenamefont {Pelc}, \citenamefont {Geisenhof},
  \citenamefont {Weitz},\ and\ \citenamefont
  {Ayuela}}]{guerrero-aviles_relative_2021}%
  \BibitemOpen
  \bibfield  {author} {\bibinfo {author} {\bibfnamefont {R.}~\bibnamefont
  {Guerrero-Avilés}}, \bibinfo {author} {\bibfnamefont {M.}~\bibnamefont
  {Pelc}}, \bibinfo {author} {\bibfnamefont {F.}~\bibnamefont {Geisenhof}},
  \bibinfo {author} {\bibfnamefont {T.}~\bibnamefont {Weitz}},\ and\ \bibinfo
  {author} {\bibfnamefont {A.}~\bibnamefont {Ayuela}},\ }\bibfield  {title}
  {\bibinfo {title} {Relative {Stability} of {Bernal} and {Rhombohedral}
  {Stackings} in {Trilayer} {Graphene} under {Distortions}},\ }\href
  {http://arxiv.org/abs/2110.06590} {\bibfield  {journal} {\bibinfo  {journal}
  {arXiv:2110.06590 [cond-mat]}\ } (\bibinfo {year} {2021})}\BibitemShut
  {NoStop}%
\bibitem [{\citenamefont {Ravnik}\ \emph {et~al.}(2019)\citenamefont {Ravnik},
  \citenamefont {Vaskivskyi}, \citenamefont {Gerasimenko}, \citenamefont
  {Diego}, \citenamefont {Vodeb}, \citenamefont {Kabanov},\ and\ \citenamefont
  {Mihailovic}}]{ravnik_strain-induced_2019}%
  \BibitemOpen
  \bibfield  {author} {\bibinfo {author} {\bibfnamefont {J.}~\bibnamefont
  {Ravnik}}, \bibinfo {author} {\bibfnamefont {I.}~\bibnamefont {Vaskivskyi}},
  \bibinfo {author} {\bibfnamefont {Y.}~\bibnamefont {Gerasimenko}}, \bibinfo
  {author} {\bibfnamefont {M.}~\bibnamefont {Diego}}, \bibinfo {author}
  {\bibfnamefont {J.}~\bibnamefont {Vodeb}}, \bibinfo {author} {\bibfnamefont
  {V.}~\bibnamefont {Kabanov}},\ and\ \bibinfo {author} {\bibfnamefont {D.~D.}\
  \bibnamefont {Mihailovic}},\ }\bibfield  {title} {\bibinfo {title}
  {Strain-induced metastable topological networks in laser-fab\-ri\-cat\-ed
  tas$_2$ polytype heterostructures for nanoscale devices},\ }\href
  {https://doi.org/10.1021/acsanm.9b00644} {\bibfield  {journal} {\bibinfo
  {journal} {ACS Applied Nano Materials}\ }\textbf {\bibinfo {volume} {2}},\
  \bibinfo {pages} {3743} (\bibinfo {year} {2019})}\BibitemShut {NoStop}%
\bibitem [{\citenamefont {Hibino}\ \emph
  {et~al.}(2009{\natexlab{b}})\citenamefont {Hibino}, \citenamefont {Mizuno},
  \citenamefont {Kageshima}, \citenamefont {Nagase},\ and\ \citenamefont
  {Yamaguchi}}]{hibino2009stacking}%
  \BibitemOpen
  \bibfield  {author} {\bibinfo {author} {\bibfnamefont {H.}~\bibnamefont
  {Hibino}}, \bibinfo {author} {\bibfnamefont {S.}~\bibnamefont {Mizuno}},
  \bibinfo {author} {\bibfnamefont {H.}~\bibnamefont {Kageshima}}, \bibinfo
  {author} {\bibfnamefont {M.}~\bibnamefont {Nagase}},\ and\ \bibinfo {author}
  {\bibfnamefont {H.}~\bibnamefont {Yamaguchi}},\ }\bibfield  {title} {\bibinfo
  {title} {{Stacking domains of epitaxial few-layer graphene on SiC(0001)}},\
  }\href {https://doi.org/10.1103/PhysRevB.80.085406} {\bibfield  {journal}
  {\bibinfo  {journal} {Physical Review B}\ }\textbf {\bibinfo {volume} {80}},\
  \bibinfo {pages} {085406} (\bibinfo {year} {2009}{\natexlab{b}})}\BibitemShut
  {NoStop}%
\bibitem [{\citenamefont {Schmidt}\ \emph {et~al.}(2011)\citenamefont
  {Schmidt}, \citenamefont {Ohta},\ and\ \citenamefont
  {Beechem}}]{schmidt_strain_2011}%
  \BibitemOpen
  \bibfield  {author} {\bibinfo {author} {\bibfnamefont {D.~A.}\ \bibnamefont
  {Schmidt}}, \bibinfo {author} {\bibfnamefont {T.}~\bibnamefont {Ohta}},\ and\
  \bibinfo {author} {\bibfnamefont {T.~E.}\ \bibnamefont {Beechem}},\
  }\bibfield  {title} {\bibinfo {title} {Strain and charge carrier coupling in
  epitaxial graphene},\ }\href {https://doi.org/10.1103/PhysRevB.84.235422}
  {\bibfield  {journal} {\bibinfo  {journal} {Physical Review B}\ }\textbf
  {\bibinfo {volume} {84}},\ \bibinfo {pages} {235422} (\bibinfo {year}
  {2011})}\BibitemShut {NoStop}%
\bibitem [{\citenamefont {Schumann}\ \emph {et~al.}(2014)\citenamefont
  {Schumann}, \citenamefont {Dubslaff}, \citenamefont {Oliveira}, \citenamefont
  {Hanke}, \citenamefont {Lopes},\ and\ \citenamefont
  {Riechert}}]{schumann2014effect}%
  \BibitemOpen
  \bibfield  {author} {\bibinfo {author} {\bibfnamefont {T.}~\bibnamefont
  {Schumann}}, \bibinfo {author} {\bibfnamefont {M.}~\bibnamefont {Dubslaff}},
  \bibinfo {author} {\bibfnamefont {M.~H.}\ \bibnamefont {Oliveira}}, \bibinfo
  {author} {\bibfnamefont {M.}~\bibnamefont {Hanke}}, \bibinfo {author}
  {\bibfnamefont {J.~M.~J.}\ \bibnamefont {Lopes}},\ and\ \bibinfo {author}
  {\bibfnamefont {H.}~\bibnamefont {Riechert}},\ }\bibfield  {title} {\bibinfo
  {title} {{Effect of buffer layer coupling on the lattice parameter of
  epitaxial graphene on SiC(0001)}},\ }\href
  {https://doi.org/10.1103/PhysRevB.90.041403} {\bibfield  {journal} {\bibinfo
  {journal} {Physical Review B}\ }\textbf {\bibinfo {volume} {90}},\ \bibinfo
  {pages} {041403} (\bibinfo {year} {2014})}\BibitemShut {NoStop}%
\bibitem [{\citenamefont {Ferralis}\ \emph {et~al.}(2008)\citenamefont
  {Ferralis}, \citenamefont {Maboudian},\ and\ \citenamefont
  {Carraro}}]{ferralis_evidence_2008}%
  \BibitemOpen
  \bibfield  {author} {\bibinfo {author} {\bibfnamefont {N.}~\bibnamefont
  {Ferralis}}, \bibinfo {author} {\bibfnamefont {R.}~\bibnamefont
  {Maboudian}},\ and\ \bibinfo {author} {\bibfnamefont {C.}~\bibnamefont
  {Carraro}},\ }\bibfield  {title} {\bibinfo {title} {Evidence of {Structural}
  {Strain} in {Epitaxial} {Graphene} {Layers} on {6H}-{SiC}(0001)},\ }\href
  {https://doi.org/10.1103/PhysRevLett.101.156801} {\bibfield  {journal}
  {\bibinfo  {journal} {Physical Review Letters}\ }\textbf {\bibinfo {volume}
  {101}},\ \bibinfo {pages} {156801} (\bibinfo {year} {2008})}\BibitemShut
  {NoStop}%
\bibitem [{\citenamefont {Bao}\ \emph {et~al.}(2016)\citenamefont {Bao},
  \citenamefont {Norimatsu}, \citenamefont {Iwata}, \citenamefont {Matsuda},
  \citenamefont {Ito},\ and\ \citenamefont {Kusunoki}}]{bao_synthesis_2016}%
  \BibitemOpen
  \bibfield  {author} {\bibinfo {author} {\bibfnamefont {J.}~\bibnamefont
  {Bao}}, \bibinfo {author} {\bibfnamefont {W.}~\bibnamefont {Norimatsu}},
  \bibinfo {author} {\bibfnamefont {H.}~\bibnamefont {Iwata}}, \bibinfo
  {author} {\bibfnamefont {K.}~\bibnamefont {Matsuda}}, \bibinfo {author}
  {\bibfnamefont {T.}~\bibnamefont {Ito}},\ and\ \bibinfo {author}
  {\bibfnamefont {M.}~\bibnamefont {Kusunoki}},\ }\bibfield  {title} {\bibinfo
  {title} {Synthesis of {Freestanding} {Graphene} on {SiC} by a
  {Rapid}-{Cooling} {Technique}},\ }\href
  {https://doi.org/10.1103/PhysRevLett.117.205501} {\bibfield  {journal}
  {\bibinfo  {journal} {Physical Review Letters}\ }\textbf {\bibinfo {volume}
  {117}},\ \bibinfo {pages} {205501} (\bibinfo {year} {2016})}\BibitemShut
  {NoStop}%
\bibitem [{\citenamefont {Lebedeva}\ and\ \citenamefont
  {Popov}(2020)}]{lebedeva_two_2020}%
  \BibitemOpen
  \bibfield  {author} {\bibinfo {author} {\bibfnamefont {I.~V.}\ \bibnamefont
  {Lebedeva}}\ and\ \bibinfo {author} {\bibfnamefont {A.~M.}\ \bibnamefont
  {Popov}},\ }\bibfield  {title} {\bibinfo {title} {Two {Phases} with
  {Different} {Domain} {Wall} {Networks} and a {Reentrant} {Phase} {Transition}
  in {Bilayer} {Graphene} under {Strain}},\ }\href
  {https://doi.org/10.1103/PhysRevLett.124.116101} {\bibfield  {journal}
  {\bibinfo  {journal} {Physical Review Letters}\ }\textbf {\bibinfo {volume}
  {124}},\ \bibinfo {pages} {116101} (\bibinfo {year} {2020})}\BibitemShut
  {NoStop}%
\bibitem [{\citenamefont {Hattab}\ \emph {et~al.}(2 08)\citenamefont {Hattab},
  \citenamefont {N’Diaye}, \citenamefont {Wall}, \citenamefont {Klein},
  \citenamefont {Jnawali}, \citenamefont {Coraux}, \citenamefont {Busse},
  \citenamefont {van Gastel}, \citenamefont {Poelsema}, \citenamefont
  {Michely}, \citenamefont {Meyer~zu Heringdorf},\ and\ \citenamefont {Horn-von
  Hoegen}}]{hattab_interplay_2012}%
  \BibitemOpen
  \bibfield  {author} {\bibinfo {author} {\bibfnamefont {H.}~\bibnamefont
  {Hattab}}, \bibinfo {author} {\bibfnamefont {A.~T.}\ \bibnamefont
  {N’Diaye}}, \bibinfo {author} {\bibfnamefont {D.}~\bibnamefont {Wall}},
  \bibinfo {author} {\bibfnamefont {C.}~\bibnamefont {Klein}}, \bibinfo
  {author} {\bibfnamefont {G.}~\bibnamefont {Jnawali}}, \bibinfo {author}
  {\bibfnamefont {J.}~\bibnamefont {Coraux}}, \bibinfo {author} {\bibfnamefont
  {C.}~\bibnamefont {Busse}}, \bibinfo {author} {\bibfnamefont
  {R.}~\bibnamefont {van Gastel}}, \bibinfo {author} {\bibfnamefont
  {B.}~\bibnamefont {Poelsema}}, \bibinfo {author} {\bibfnamefont
  {T.}~\bibnamefont {Michely}}, \bibinfo {author} {\bibfnamefont {F.-J.}\
  \bibnamefont {Meyer~zu Heringdorf}},\ and\ \bibinfo {author} {\bibfnamefont
  {M.}~\bibnamefont {Horn-von Hoegen}},\ }\bibfield  {title} {\bibinfo {title}
  {Interplay of wrinkles, strain, and lattice parameter in graphene on
  iridium},\ }\href {https://doi.org/10.1021/nl203530t} {\bibfield  {journal}
  {\bibinfo  {journal} {Nano Letters}\ }\textbf {\bibinfo {volume} {12}},\
  \bibinfo {pages} {678} (\bibinfo {year} {2012-08})}\BibitemShut {NoStop}%
\bibitem [{\citenamefont {Sutter}\ \emph {et~al.}(2009)\citenamefont {Sutter},
  \citenamefont {Sadowski},\ and\ \citenamefont
  {Sutter}}]{sutter_graphene_2009}%
  \BibitemOpen
  \bibfield  {author} {\bibinfo {author} {\bibfnamefont {P.}~\bibnamefont
  {Sutter}}, \bibinfo {author} {\bibfnamefont {J.~T.}\ \bibnamefont
  {Sadowski}},\ and\ \bibinfo {author} {\bibfnamefont {E.}~\bibnamefont
  {Sutter}},\ }\bibfield  {title} {\bibinfo {title} {Graphene on {Pt}(111):
  {Growth} and substrate interaction},\ }\href
  {https://doi.org/10.1103/PhysRevB.80.245411} {\bibfield  {journal} {\bibinfo
  {journal} {Physical Review B}\ }\textbf {\bibinfo {volume} {80}},\ \bibinfo
  {pages} {245411} (\bibinfo {year} {2009})}\BibitemShut {NoStop}%
\bibitem [{\citenamefont {Benschop}\ \emph {et~al.}(2021)\citenamefont
  {Benschop}, \citenamefont {de~Jong}, \citenamefont {Stepanov}, \citenamefont
  {Lu}, \citenamefont {Stalman}, \citenamefont {van~der Molen}, \citenamefont
  {Efetov},\ and\ \citenamefont {Allan}}]{benschop_measuring_2021}%
  \BibitemOpen
  \bibfield  {author} {\bibinfo {author} {\bibfnamefont {T.}~\bibnamefont
  {Benschop}}, \bibinfo {author} {\bibfnamefont {T.~A.}\ \bibnamefont
  {de~Jong}}, \bibinfo {author} {\bibfnamefont {P.}~\bibnamefont {Stepanov}},
  \bibinfo {author} {\bibfnamefont {X.}~\bibnamefont {Lu}}, \bibinfo {author}
  {\bibfnamefont {V.}~\bibnamefont {Stalman}}, \bibinfo {author} {\bibfnamefont
  {S.~J.}\ \bibnamefont {van~der Molen}}, \bibinfo {author} {\bibfnamefont
  {D.~K.}\ \bibnamefont {Efetov}},\ and\ \bibinfo {author} {\bibfnamefont
  {M.~P.}\ \bibnamefont {Allan}},\ }\bibfield  {title} {\bibinfo {title}
  {Measuring local moiré lattice heterogeneity of twisted bilayer graphene},\
  }\href {https://doi.org/10.1103/PhysRevResearch.3.013153} {\bibfield
  {journal} {\bibinfo  {journal} {Physical Review Research}\ }\textbf {\bibinfo
  {volume} {3}},\ \bibinfo {pages} {013153} (\bibinfo {year}
  {2021})}\BibitemShut {NoStop}%
\bibitem [{\citenamefont {Gutiérrez}\ \emph {et~al.}(2016)\citenamefont
  {Gutiérrez}, \citenamefont {Kim}, \citenamefont {Brown}, \citenamefont
  {Schiros}, \citenamefont {Nordlund}, \citenamefont {Lochocki}, \citenamefont
  {Shen}, \citenamefont {Park},\ and\ \citenamefont
  {Pasupathy}}]{gutierrez_imaging_2016}%
  \BibitemOpen
  \bibfield  {author} {\bibinfo {author} {\bibfnamefont {C.}~\bibnamefont
  {Gutiérrez}}, \bibinfo {author} {\bibfnamefont {C.-J.}\ \bibnamefont {Kim}},
  \bibinfo {author} {\bibfnamefont {L.}~\bibnamefont {Brown}}, \bibinfo
  {author} {\bibfnamefont {T.}~\bibnamefont {Schiros}}, \bibinfo {author}
  {\bibfnamefont {D.}~\bibnamefont {Nordlund}}, \bibinfo {author}
  {\bibfnamefont {E.~B.}\ \bibnamefont {Lochocki}}, \bibinfo {author}
  {\bibfnamefont {K.~M.}\ \bibnamefont {Shen}}, \bibinfo {author}
  {\bibfnamefont {J.}~\bibnamefont {Park}},\ and\ \bibinfo {author}
  {\bibfnamefont {A.~N.}\ \bibnamefont {Pasupathy}},\ }\bibfield  {title}
  {\bibinfo {title} {Imaging chiral symmetry breaking from {Kekulé} bond order
  in graphene},\ }\href {https://doi.org/10.1038/nphys3776} {\bibfield
  {journal} {\bibinfo  {journal} {Nature Physics}\ }\textbf {\bibinfo {volume}
  {12}},\ \bibinfo {pages} {950} (\bibinfo {year} {2016})}\BibitemShut
  {NoStop}%
\bibitem [{\citenamefont {Shao}\ \emph {et~al.}(2013)\citenamefont {Shao},
  \citenamefont {Wang}, \citenamefont {Misra},\ and\ \citenamefont
  {Hoagland}}]{shao_spiral_2013}%
  \BibitemOpen
  \bibfield  {author} {\bibinfo {author} {\bibfnamefont {S.}~\bibnamefont
  {Shao}}, \bibinfo {author} {\bibfnamefont {J.}~\bibnamefont {Wang}}, \bibinfo
  {author} {\bibfnamefont {A.}~\bibnamefont {Misra}},\ and\ \bibinfo {author}
  {\bibfnamefont {R.~G.}\ \bibnamefont {Hoagland}},\ }\bibfield  {title}
  {\bibinfo {title} {Spiral {Patterns} of {Dislocations} at {Nodes} in (111)
  {Semi}-coherent {FCC} {Interfaces}},\ }\href
  {https://doi.org/10.1038/srep02448} {\bibfield  {journal} {\bibinfo
  {journal} {Scientific Reports}\ }\textbf {\bibinfo {volume} {3}},\ \bibinfo
  {pages} {2448} (\bibinfo {year} {2013})}\BibitemShut {NoStop}%
\bibitem [{\citenamefont {Hamilton}\ and\ \citenamefont
  {Foiles}(1995)}]{hamilton_misfit_1995}%
  \BibitemOpen
  \bibfield  {author} {\bibinfo {author} {\bibfnamefont {J.~C.}\ \bibnamefont
  {Hamilton}}\ and\ \bibinfo {author} {\bibfnamefont {S.~M.}\ \bibnamefont
  {Foiles}},\ }\bibfield  {title} {\bibinfo {title} {Misfit {Dislocation}
  {Structure} for {Close}-{Packed} {Metal}-{Metal} {Interfaces}},\ }\href
  {https://doi.org/10.1103/PhysRevLett.75.882} {\bibfield  {journal} {\bibinfo
  {journal} {Physical Review Letters}\ }\textbf {\bibinfo {volume} {75}},\
  \bibinfo {pages} {882} (\bibinfo {year} {1995})}\BibitemShut {NoStop}%
\bibitem [{\citenamefont {Günther}\ \emph {et~al.}(1995)\citenamefont
  {Günther}, \citenamefont {Vrijmoeth}, \citenamefont {Hwang},\ and\
  \citenamefont {Behm}}]{gunther_strain_1995}%
  \BibitemOpen
  \bibfield  {author} {\bibinfo {author} {\bibfnamefont {C.}~\bibnamefont
  {Günther}}, \bibinfo {author} {\bibfnamefont {J.}~\bibnamefont {Vrijmoeth}},
  \bibinfo {author} {\bibfnamefont {R.~Q.}\ \bibnamefont {Hwang}},\ and\
  \bibinfo {author} {\bibfnamefont {R.~J.}\ \bibnamefont {Behm}},\ }\bibfield
  {title} {\bibinfo {title} {Strain relaxation in hexagonally close-packed
  metal-metal interfaces},\ }\href {https://doi.org/10.1103/PhysRevLett.74.754}
  {\bibfield  {journal} {\bibinfo  {journal} {Physical Review Letters}\
  }\textbf {\bibinfo {volume} {74}},\ \bibinfo {pages} {754} (\bibinfo {year}
  {1995})}\BibitemShut {NoStop}%
\bibitem [{\citenamefont {Snyman}\ and\ \citenamefont
  {Snyman}(1981)}]{snyman_computed_1981}%
  \BibitemOpen
  \bibfield  {author} {\bibinfo {author} {\bibfnamefont {J.~A.}\ \bibnamefont
  {Snyman}}\ and\ \bibinfo {author} {\bibfnamefont {H.~C.}\ \bibnamefont
  {Snyman}},\ }\bibfield  {title} {\bibinfo {title} {Computed epitaxial
  monolayer structures: {III}. {Two}-dimensional model: zero average strain
  monolayer structures in the case of hexagonal interfacial symmetry},\ }\href
  {https://doi.org/10.1016/0039-6028(81)90168-0} {\bibfield  {journal}
  {\bibinfo  {journal} {Surface Science}\ }\textbf {\bibinfo {volume} {105}},\
  \bibinfo {pages} {357} (\bibinfo {year} {1981})}\BibitemShut {NoStop}%
\bibitem [{\citenamefont {Quan}\ \emph {et~al.}(2018)\citenamefont {Quan},
  \citenamefont {He},\ and\ \citenamefont {Ni}}]{quan_tunable_2018}%
  \BibitemOpen
  \bibfield  {author} {\bibinfo {author} {\bibfnamefont {S.}~\bibnamefont
  {Quan}}, \bibinfo {author} {\bibfnamefont {L.}~\bibnamefont {He}},\ and\
  \bibinfo {author} {\bibfnamefont {Y.}~\bibnamefont {Ni}},\ }\bibfield
  {title} {\bibinfo {title} {Tunable mosaic structures in van der {Waals}
  layered materials},\ }\href {https://doi.org/10.1039/C8CP04360D} {\bibfield
  {journal} {\bibinfo  {journal} {Physical Chemistry Chemical Physics}\
  }\textbf {\bibinfo {volume} {20}},\ \bibinfo {pages} {25428} (\bibinfo {year}
  {2018})}\BibitemShut {NoStop}%
\bibitem [{\citenamefont {Yoo}\ \emph {et~al.}(2019)\citenamefont {Yoo},
  \citenamefont {Engelke}, \citenamefont {Carr}, \citenamefont {Fang},
  \citenamefont {Zhang}, \citenamefont {Cazeaux}, \citenamefont {Sung},
  \citenamefont {Hovden}, \citenamefont {Tsen}, \citenamefont {Taniguchi},
  \citenamefont {Watanabe}, \citenamefont {Yi}, \citenamefont {Kim},
  \citenamefont {Luskin}, \citenamefont {Tadmor}, \citenamefont {Kaxiras},\
  and\ \citenamefont {Kim}}]{Yoo2019}%
  \BibitemOpen
  \bibfield  {author} {\bibinfo {author} {\bibfnamefont {H.}~\bibnamefont
  {Yoo}}, \bibinfo {author} {\bibfnamefont {R.}~\bibnamefont {Engelke}},
  \bibinfo {author} {\bibfnamefont {S.}~\bibnamefont {Carr}}, \bibinfo {author}
  {\bibfnamefont {S.}~\bibnamefont {Fang}}, \bibinfo {author} {\bibfnamefont
  {K.}~\bibnamefont {Zhang}}, \bibinfo {author} {\bibfnamefont
  {P.}~\bibnamefont {Cazeaux}}, \bibinfo {author} {\bibfnamefont {S.~H.}\
  \bibnamefont {Sung}}, \bibinfo {author} {\bibfnamefont {R.}~\bibnamefont
  {Hovden}}, \bibinfo {author} {\bibfnamefont {A.~W.}\ \bibnamefont {Tsen}},
  \bibinfo {author} {\bibfnamefont {T.}~\bibnamefont {Taniguchi}}, \bibinfo
  {author} {\bibfnamefont {K.}~\bibnamefont {Watanabe}}, \bibinfo {author}
  {\bibfnamefont {G.-C.}\ \bibnamefont {Yi}}, \bibinfo {author} {\bibfnamefont
  {M.}~\bibnamefont {Kim}}, \bibinfo {author} {\bibfnamefont {M.}~\bibnamefont
  {Luskin}}, \bibinfo {author} {\bibfnamefont {E.~B.}\ \bibnamefont {Tadmor}},
  \bibinfo {author} {\bibfnamefont {E.}~\bibnamefont {Kaxiras}},\ and\ \bibinfo
  {author} {\bibfnamefont {P.}~\bibnamefont {Kim}},\ }\bibfield  {title}
  {\bibinfo {title} {Atomic and electronic reconstruction at the van der waals
  interface in twisted bilayer graphene},\ }\href
  {https://doi.org/10.1038/s41563-019-0346-z} {\bibfield  {journal} {\bibinfo
  {journal} {Nature Materials}\ }\textbf {\bibinfo {volume} {18}},\ \bibinfo
  {pages} {448} (\bibinfo {year} {2019})}\BibitemShut {NoStop}%
\bibitem [{\citenamefont {Verbakel}\ \emph {et~al.}(2021)\citenamefont
  {Verbakel}, \citenamefont {Yao}, \citenamefont {Sotthewes},\ and\
  \citenamefont {Zandvliet}}]{verbakel_valley-protected_2021}%
  \BibitemOpen
  \bibfield  {author} {\bibinfo {author} {\bibfnamefont {J.~D.}\ \bibnamefont
  {Verbakel}}, \bibinfo {author} {\bibfnamefont {Q.}~\bibnamefont {Yao}},
  \bibinfo {author} {\bibfnamefont {K.}~\bibnamefont {Sotthewes}},\ and\
  \bibinfo {author} {\bibfnamefont {H.~J.~W.}\ \bibnamefont {Zandvliet}},\
  }\bibfield  {title} {\bibinfo {title} {Valley-protected one-dimensional
  states in small-angle twisted bilayer graphene},\ }\href
  {https://doi.org/10.1103/PhysRevB.103.165134} {\bibfield  {journal} {\bibinfo
   {journal} {Physical Review B}\ }\textbf {\bibinfo {volume} {103}},\ \bibinfo
  {pages} {165134} (\bibinfo {year} {2021})}\BibitemShut {NoStop}%
\bibitem [{Note1()}]{Note1}%
  \BibitemOpen
  \bibinfo {note} {For any realistic atomic strain, this expression is valid
  for $|\theta _a| \lesssim \epsilon $, but the crossover at $\theta = 30^\circ
  $ makes it impossible to distinguish the sign for $|\theta _a| \gtrsim
  \protect \frac {\epsilon }{2}$ in BF-LEEM.}\BibitemShut {Stop}%
\bibitem [{\citenamefont {Kerelsky}\ \emph {et~al.}(2019)\citenamefont
  {Kerelsky}, \citenamefont {McGilly}, \citenamefont {Kennes}, \citenamefont
  {Xian}, \citenamefont {Yankowitz}, \citenamefont {Chen}, \citenamefont
  {Watanabe}, \citenamefont {Taniguchi}, \citenamefont {Hone}, \citenamefont
  {Dean}, \citenamefont {Rubio},\ and\ \citenamefont
  {Pasupathy}}]{Kerelsky2019}%
  \BibitemOpen
  \bibfield  {author} {\bibinfo {author} {\bibfnamefont {A.}~\bibnamefont
  {Kerelsky}}, \bibinfo {author} {\bibfnamefont {L.~J.}\ \bibnamefont
  {McGilly}}, \bibinfo {author} {\bibfnamefont {D.~M.}\ \bibnamefont {Kennes}},
  \bibinfo {author} {\bibfnamefont {L.}~\bibnamefont {Xian}}, \bibinfo {author}
  {\bibfnamefont {M.}~\bibnamefont {Yankowitz}}, \bibinfo {author}
  {\bibfnamefont {S.}~\bibnamefont {Chen}}, \bibinfo {author} {\bibfnamefont
  {K.}~\bibnamefont {Watanabe}}, \bibinfo {author} {\bibfnamefont
  {T.}~\bibnamefont {Taniguchi}}, \bibinfo {author} {\bibfnamefont
  {J.}~\bibnamefont {Hone}}, \bibinfo {author} {\bibfnamefont {C.}~\bibnamefont
  {Dean}}, \bibinfo {author} {\bibfnamefont {A.}~\bibnamefont {Rubio}},\ and\
  \bibinfo {author} {\bibfnamefont {A.~N.}\ \bibnamefont {Pasupathy}},\
  }\bibfield  {title} {\bibinfo {title} {Maximized electron interactions at the
  magic angle in twisted bilayer graphene},\ }\href
  {https://doi.org/10.1038/s41586-019-1431-9} {\bibfield  {journal} {\bibinfo
  {journal} {Nature}\ }\textbf {\bibinfo {volume} {572}},\ \bibinfo {pages}
  {95} (\bibinfo {year} {2019})}\BibitemShut {NoStop}%
\bibitem [{\citenamefont {Kimoto}\ and\ \citenamefont
  {Watanabe}(2020)}]{kimoto_defect_2020}%
  \BibitemOpen
  \bibfield  {author} {\bibinfo {author} {\bibfnamefont {T.}~\bibnamefont
  {Kimoto}}\ and\ \bibinfo {author} {\bibfnamefont {H.}~\bibnamefont
  {Watanabe}},\ }\bibfield  {title} {\bibinfo {title} {Defect engineering in
  {SiC} technology for high-voltage power devices},\ }\href
  {https://doi.org/10.35848/1882-0786/abc787} {\bibfield  {journal} {\bibinfo
  {journal} {Applied Physics Express}\ }\textbf {\bibinfo {volume} {13}},\
  \bibinfo {pages} {120101} (\bibinfo {year} {2020})}\BibitemShut {NoStop}%
\bibitem [{\citenamefont {Łażewski}\ \emph {et~al.}(2019)\citenamefont
  {Łażewski}, \citenamefont {Jochym}, \citenamefont {Piekarz}, \citenamefont
  {Sternik}, \citenamefont {Parlinski}, \citenamefont {Cholewiński},
  \citenamefont {Dłużewski},\ and\ \citenamefont
  {Krukowski}}]{lazewski_dft_2019}%
  \BibitemOpen
  \bibfield  {author} {\bibinfo {author} {\bibfnamefont {J.}~\bibnamefont
  {Łażewski}}, \bibinfo {author} {\bibfnamefont {P.~T.}\ \bibnamefont
  {Jochym}}, \bibinfo {author} {\bibfnamefont {P.}~\bibnamefont {Piekarz}},
  \bibinfo {author} {\bibfnamefont {M.}~\bibnamefont {Sternik}}, \bibinfo
  {author} {\bibfnamefont {K.}~\bibnamefont {Parlinski}}, \bibinfo {author}
  {\bibfnamefont {J.}~\bibnamefont {Cholewiński}}, \bibinfo {author}
  {\bibfnamefont {P.}~\bibnamefont {Dłużewski}},\ and\ \bibinfo {author}
  {\bibfnamefont {S.}~\bibnamefont {Krukowski}},\ }\bibfield  {title} {\bibinfo
  {title} {{DFT} modelling of the edge dislocation in {4H}-{SiC}},\ }\href
  {https://doi.org/10.1007/s10853-019-03630-5} {\bibfield  {journal} {\bibinfo
  {journal} {Journal of Materials Science}\ }\textbf {\bibinfo {volume} {54}},\
  \bibinfo {pages} {10737} (\bibinfo {year} {2019})}\BibitemShut {NoStop}%
\end{thebibliography}%

\end{document}